\documentclass[a4paper,12pt]{article}

\pdfoutput=1 

\usepackage[hmargin=.71in,vmargin=1.1in]{geometry}
\usepackage{indentfirst}
\usepackage{amsfonts}
\usepackage{mathrsfs}
\usepackage{amsmath}
\usepackage{mathabx}
\usepackage{amssymb}
\usepackage{authblk}
\usepackage{cite}
\usepackage{xcolor}
\usepackage{mathtools}
\usepackage{tensor}
\usepackage{physics}
\usepackage{graphicx}
\usepackage{bm}
\usepackage{upgreek}
\usepackage{braket}
\usepackage{color,soul}
\usepackage{csquotes}
\usepackage{caption}
\usepackage{subcaption}
\usepackage{slashed}
\usepackage{mathtools}
\usepackage{booktabs}

\usepackage[bookmarksnumbered=true,bookmarksopen=true]{hyperref}
 \hypersetup{colorlinks,
             linkcolor=[rgb]{0,0.3,0.6}, 
             citecolor=[rgb]{0,0.3,0.6}, 
             urlcolor=[rgb]{0,0.3,0.6}}

\newcommand\mi{\mathrm{i}}
\newcommand\me{\mathrm{e}}

\newcommand\pp{\uppi}
\newcommand{\dif}{\mathrm{d}}

\usepackage{tikz}
\usetikzlibrary{matrix}

\begin{document}

\title{\Large\textbf{Branch-dependent ringdown in black-bounce spacetimes: imprints of matter-source ambiguity on quasinormal modes}}

\author[a]{Hao Yang\thanks{hyang@ucas.ac.cn}}
\author[b]{Chen Lan\thanks{stlanchen@126.com}}

\affil[a]{\normalsize{\em School of Fundamental Physics and Mathematical Sciences, Hangzhou Institute for Advanced Study, UCAS, Hangzhou 310024, China}}
\affil[b]{\normalsize{\em Department of Physics, Yantai University, 30 Qingquan Road, Yantai 264005, China}}

\date{ }

\maketitle

\begin{abstract}

Regular black holes and black-bounce spacetimes frequently emerge in theoretical frameworks beyond general relativity, as well as in general relativity coupled to non-linear sources. A profound complication in these frameworks is source ambiguity: a single spacetime metric can often be supported by multiple, inequivalent matter-source interpretations, such as an anisotropic fluid or nonlinear electrodynamics (NED) coupled to a scalar field. We investigate how this fundamental degeneracy dynamically imprints on axial gravitational perturbations within the Simpson-Visser spacetime, which smoothly transitions from a regular black hole (BH) to a traversable wormhole (WH) at a critical bounce parameter $a=2M$. By deriving the exact master equations for each interpretation, we perform time-domain numerical evolutions to extract the quasinormal modes (QNMs) via Prony fitting. In the BH branch ($a\le2M$), the NED interpretation exhibits faster QNM damping than the fluid model, driven by enhanced energy leakage through the coupled electromagnetic channel alongside horizon absorption. Conversely, in the WH branch ($a>2M$), the NED coupled system produces longer-lived fundamental modes. This reduced damping is governed by subradiant-like interference that actively suppresses radiative losses to the two asymptotically flat regions. This branch-dependent dynamics, analogous to decay-width redistribution in open non-Hermitian quantum systems, demonstrates that matter-source ambiguity leaves distinct, observable signatures in ringdown waveforms. Our findings establish that gravitational-wave spectroscopy can systematically break the degeneracy of source interpretations, providing a novel empirical pathway to probe the physical nature of exotic compact objects.

\end{abstract}

\tableofcontents

\newpage

\section{Introduction}
\label{sec:intro}

Regular black holes (RBHs) are spacetime geometries free of curvature singularities, with the Bardeen black hole serving as an early prototype~\cite{Bardeen:1968nsg}. This concept, however, stands in sharp tension with the Hawking--Penrose singularity theorems: within general relativity, gravitational collapse generically culminates in spacetime singularities provided standard energy conditions hold~\cite{Hawking:1973uf,Wald:1984rg}. Consequently, realizing an RBH within an effective general-relativistic framework necessitates a supporting matter sector that violates at least some of these energy conditions, differing qualitatively from ordinary classical matter~\cite{Curiel:2014zba,Balart:2014jia}. This paradigm naturally raises two central questions: (i) what specific matter fields can consistently source a given regular geometry, and (ii) how can such exotic matter sectors be distinguished observationally?

Since the turn of the century, a broad spectrum of RBH models has been developed, motivated by either effective matter sources or modified gravitational dynamics~\cite{Lan:2023cvz,Junior:2023qaq,Bonanno:2023rzk,Bueno:2024eig,Bueno:2024dgm,Bueno:2024zsx,Bueno:2025zaj}. Notable examples include noncommutative-inspired constructions~\cite{Nicolini:2005vd}, solutions sourced by nonlinear electrodynamics (NED)~\cite{Ayon-Beato:2000mjt}, and quantum-gravity-motivated effective cores~\cite{Modesto:2008im}. A profound complication in this landscape is the inherent degeneracy of the field equations: the physical interpretation of the source supporting a given metric is not necessarily unique. A highly illustrative example is the black-bounce framework, which smoothly interpolates between regular black holes and traversable wormholes depending on parameter choices~\cite{Simpson:2018tsi,Franzin:2021vnj,Lobo:2020ffi}. For these black-bounce spacetimes, several inequivalent realizations of the supporting matter sector have been rigorously constructed, ranging from anisotropic-fluid descriptions to nonlinear electrodynamics in the presence of scalar fields~\cite{Rodrigues:2023vtm,Alencar:2024yvh,Cordeiro:2025ivw,Lessa:2024erf,Junior:2025sjr}. This source ambiguity strongly motivates a systematic investigation into whether different matter-field interpretations of the identical background geometry can leave distinct observational imprints.

Gravitational-wave ringdown provides an ideal testing ground for breaking this degeneracy. The ringdown waveform is dominated by quasinormal modes (QNMs), whose complex frequencies directly encode the dissipative response of the remnant compact object~\cite{Kokkotas:1999bd,Berti:2009kk,Konoplya:2011qq,Duran-Cabaces:2025sly,Franzin:2023slm,Tian:2025uvk}. Traditionally, studies of RBH ringdown focus primarily on the linearized perturbations of the spacetime metric itself, typically decomposed into axial and polar sectors following the classic black-hole perturbation formalism~\cite{chandrasekhar1998mathematical,Konoplya:2023aph,Stashko:2024wuq,del-Corral:2022kbk}. However, for non-vacuum spacetimes supported by non-standard matter, a fully consistent perturbative treatment must concurrently track the fluctuations of the supporting fields\cite{Chen:2019iuo}. These matter perturbations generally couple nontrivially to the gravitational sector, fundamentally altering the governing dynamics and, consequently, the observable ringdown signal.

In this work, we investigate the axial gravitational perturbations of the Simpson-Visser (SV) black-bounce spacetime under structurally different matter-source interpretations. By deriving the exact master equations for each interpretation, we evolve the perturbations numerically in the time domain and extract the resulting QNMs using Prony fitting. By comparing the black-hole branch (characterized by horizon absorption) with the wormhole branch (featuring two asymptotically flat regions but lacking a horizon), we systematically quantify how matter-source ambiguity is dynamically imprinted on the axial ringdown. Notably, we demonstrate that while the anisotropic-fluid model yields a standard single-channel wave equation, the nonlinear-electrodynamics-plus-scalar interpretation generates a genuinely coupled two-channel system. This coupling facilitates a redistribution of effective decay widths between the eigenmodes—a hallmark of open, non-Hermitian radiative systems~\cite{Rotter2009,GrossHaroche1982}—ultimately producing branch-dependent damping behaviors that are entirely absent in the uncoupled fluid description.

The paper is organized as follows. In Sec.~\ref{sec:Spacetime}, we review the black-bounce background geometry alongside three representative matter-source interpretations. Section~\ref{sec:Derivation} details the derivation of the axial perturbation equations for each of these interpretations. In Sec.~\ref{sec:InBound}, we specify the initial and boundary conditions, as well as the numerical methodology employed for time-domain evolution and QNM extraction. Section~\ref{sec:Dynamical} presents our main results, comparing the ringdown dynamics and QNM spectra across both the matter interpretations and the geometric branches. Finally, we conclude in Sec.~\ref{sec:Conclusion} with a discussion of the physical implications and future observational prospects. All numerically computed QNM frequencies are provided in the supplementary material for verification.

\section{Matter sources in black-bounce spacetime}
\label{sec:Spacetime}

In this paper, we consider the static, spherically symmetric black-bounce spacetime described by the following metric~\cite{Rodrigues:2023vtm,Alencar:2024yvh}
\begin{equation}\label{eq:general-metric}
    \dif s^2 = g_{\mu\nu}\dif x^\mu \dif x^\nu = f(r)\dif t^2 - f(r)^{-1} \dif r^2 - \Sigma^2(r)\left(\dif\theta^2 + \sin^2\theta\dif \varphi^2\right),
\end{equation}
where $f(r)$ and $\Sigma(r)$ are strictly functions of the radial coordinate $r$, which spans the range $(-\infty, +\infty)$. A representative example of such a geometry is the SV solution~\cite{Simpson:2018tsi},
\begin{equation}
\label{eq:par_sv}
    f(r) = 1 - \frac{2M}{\sqrt{r^2+a^2}}, \quad \text{and}\quad \Sigma(r) = \sqrt{r^2+a^2},
\end{equation}
where $M$ is the mass parameter and $a$ is the bounce parameter. 

The SV solution perfectly encapsulates two defining features of black-bounce spacetimes. First, the domains $r>0$ and $r<0$ represent two symmetric universes connected at $r=0$ by a regular throat, avoiding any spacetime singularity. Second, the causal structure of the spacetime depends critically on the value of the parameter $a$. For $a \le 2M$, the spacetime possesses an event horizon and describes a RBH. Conversely, for $a > 2M$, the event horizon is absent, rendering the geometry a traversable wormhole (WH).

From the Einstein field equations, the nonzero components of the energy-momentum tensor for the black-bounce background are evaluated as
\begin{subequations}\label{eq:T}
    \begin{equation}\label{eq:T00}
        \kappa^2{T^0}_0 = {G^0}_0 = -\frac{f'(r)\Sigma'(r)}{\Sigma(r)} - \frac{f(r)\Sigma'(r)^2}{\Sigma(r)^2} - \frac{2f(r)\Sigma''(r)}{\Sigma(r)} + \frac{1}{\Sigma(r)^2},
    \end{equation}
    \begin{equation}
        \kappa^2{T^1}_1 = {G^1}_1 = -\frac{f'(r)\Sigma'(r)}{\Sigma(r)} - \frac{f(r)\Sigma'(r)^2}{\Sigma(r)^2} + \frac{1}{\Sigma(r)^2},
    \end{equation}
    \begin{equation}
        \kappa^2 {T^2}_2 = \kappa^2 {T^3}_3 = {G^2}_2 = {G^3}_3 = -\frac{f'(r)\Sigma'(r)}{\Sigma(r)} - \frac{f''(r)}{2} - \frac{f(r)\Sigma''(r)}{\Sigma(r)},
    \end{equation}    
\end{subequations}
where ${T^{\mu}}_{\nu}$ is the energy-momentum tensor, ${G^\mu}_\nu$ is the Einstein tensor, $\kappa^2=8\pp$, and primes denote derivatives with respect to $r$. The non-vanishing nature of these components indicates the presence of an underlying matter distribution that sources the black-bounce geometry. 

However, the physical interpretation of this matter source is inherently non-unique. In this study, we focus on three distinct interpretations: an anisotropic fluid (AF), NED coupled to a scalar field in a static electric configuration, and NED coupled to a scalar field in a static magnetic configuration. It is worth noting that the SV geometry violates all classical energy conditions, as established in Ref.~\cite{Simpson:2018tsi} and further generalized in Ref.~\cite{Lobo:2020ffi}. Because the macroscopic energy-momentum tensor is completely dictated by the purely geometric sector (the Einstein tensor) of the field equations, this violation is identical across all three matter interpretations. Within the framework of general relativity, these fundamentally distinct physical models yield the exact same effective energy-momentum tensor, leading to identical energy condition violations. We detail the specifics of each interpretation in the following subsections.

\subsection{Anisotropic fluid model}
\label{subsec:Afi}

From a phenomenological perspective, the field source of the black-bounce spacetime can be modeled as an AF. The corresponding energy-momentum tensor for this fluid is expressed as~\cite{Chen:2019iuo}
\begin{equation}\label{eq:TAF}
    T^{\mathrm{AF}}_{\mu\nu} = (\rho+p_2)u_\mu u_\nu + (p_1-p_2)x_\mu x_\nu - p_2 g_{\mu\nu},
\end{equation}
where $\rho$, $p_1$, and $p_2$ are the energy density, radial pressure, and tangential pressure, respectively. Here, $u^\mu$ is the unit timelike four-velocity, and $x^\mu$ is a unit spacelike vector orthogonal to both $u^\mu$ and the angular directions. Therefore, $u^\mu$ and $x^\mu$ satisfy the constraints
\begin{equation}\label{eq:umu-xmu}
    u_\mu u^\mu = 1,\quad x_\mu x^\mu = -1,\quad u_\mu x^\mu = 0,
\end{equation}
and can be parameterized as
\begin{equation}\label{eq:umu-xmu-2}
    u_\mu = \left(u_t, 0, 0, 0\right), \quad x_\mu = \left(0, x_r, 0, 0\right).
\end{equation}
The non-zero components of the fluid's energy-momentum tensor are directly related to the Einstein tensor via
\begin{subequations}\label{eq:nonzeroTAF}
    \begin{equation}
        T^{\mathrm{AF}}_{00} = \rho g_{00},\quad \rho = \frac{G^0_0}{\kappa^2},
    \end{equation}
    \begin{equation}
        T^{\mathrm{AF}}_{11} = -p_1 g_{11},\quad p_1 = -\frac{G^1_1}{\kappa^2},
    \end{equation}
    \begin{equation}
        T^{\mathrm{AF}}_{22} = -p_2 g_{22},\quad T^{\mathrm{AF}}_{33} = -p_2 g_{33},\quad p_2 = -\frac{G^2_2}{\kappa^2}.
    \end{equation}
\end{subequations}

\subsection{Nonlinear electrodynamics coupled to a scalar field}

Alternatively, the source can be interpreted via a Hilbert-Einstein action coupling general relativity to NED and a scalar field
\begin{equation}\label{eq:action}
    S = \int \sqrt{-g}\ \dif^4 x \left[\frac{R}{2\kappa^2} + \epsilon g^{\mu\nu}\partial_\mu\phi\partial_\nu\phi - V(\phi) - L(F)\right],
\end{equation}
where $g$ is the determinant of the metric, $R$ is the Ricci scalar, $\phi$ is the scalar field, $V(\phi)$ is the scalar potential, and the nonlinear electromagnetic Lagrangian $L(F)$ is a function of the electromagnetic invariant $F = \frac{1}{4} F_{\mu\nu}F^{\mu\nu}$. The parameter $\epsilon = \pm 1$ depends on whether the scalar field is canonical $(+)$ or phantom $(-)$.

Varying this action yields the following field equations:
\begin{equation}\label{eq:Einstein-eq}
    G_{\mu\nu} = R_{\mu\nu} - \frac{1}{2}g_{\mu\nu}R = \kappa^2\left(T^{\phi}_{\mu\nu} + T^{\mathrm{EM}}_{\mu\nu}\right),
\end{equation}
\begin{equation}\label{eq:KG-eq}
    2\epsilon\nabla_\mu\nabla^\mu\phi = -\frac{\dif V(\phi)}{\dif\phi},
\end{equation}
\begin{equation}\label{eq:maxwell-eq}
    \nabla_\mu\left(L_F F^{\mu\nu}\right) = 0,
\end{equation}
where $L_F = \dif L/\dif F$. The energy-momentum tensors for the electromagnetic field ($T^{\mathrm{EM}}_{\mu\nu}$) and the scalar field ($T^{\phi}_{\mu\nu}$) are expressed as follows 
\begin{equation}\label{eq:TEM}
    T^{\mathrm{EM}}_{\mu\nu} = g_{\mu\nu}L(F) - L_F{F_\nu}^\alpha F_{\mu\alpha},
\end{equation}
\begin{equation}\label{eq:Tphi}
    T^\phi_{\mu\nu} = 2\epsilon\partial_\nu\phi\partial_\mu\phi - g_{\mu\nu}\left(\epsilon\partial^\alpha\phi\partial_\alpha\phi - V(\phi)\right).
\end{equation}

\subsubsection{Static electric configurations}

For a static electric field, the only nonzero components of the electromagnetic field tensor are $F^{01} = -F^{10}$. Using Eqs.~\eqref{eq:general-metric} and~\eqref{eq:maxwell-eq}, the nonlinear Maxwell equation simplifies to
\begin{equation}
    \frac{1}{\sqrt{-g}}\frac{\partial}{\partial r}\left(\sqrt{-g} L^{(E)}_F F^{10}\right) = 0,
\end{equation}
where the superscript $(E)$ denotes the static electric configuration. Integrating this equation allows us to define the electric charge $q_{e}$ as an integration constant
\begin{equation}
   \Sigma^2(r)L^{(E)}_F F^{10} = \text{constant} = q_{e}.
\end{equation}
According to Eq.~\eqref{eq:TEM}, the stress-energy tensor of the static electric field is given by
\begin{equation}\label{eq:TSE}
    {{T^{(E)}}^\mu}_\nu = \text{diag}\left(L^{(E)}(r) + \frac{q_e^2}{L^{(E)}_F(r)\Sigma^4(r)}, L^{(E)}(r) + \frac{q_e^2}{L^{(E)}_F(r)\Sigma^4(r)}, L^{(E)}(r), L^{(E)}(r)\right).
\end{equation}
Solving the coupled system of Eqs.~\eqref{eq:Einstein-eq}, \eqref{eq:KG-eq}, \eqref{eq:maxwell-eq}, \eqref{eq:Tphi}, and~\eqref{eq:TSE} uniquely determines the forms of $L^{(E)}(r)$, $L^{(E)}_F(r)$, $\phi(r)$, and $V(\phi)$. Of particular importance for the subsequent perturbation analysis is $L^{(E)}_F(r)$, which takes the exact form
\begin{equation}\label{eq:LFE}
    L^{(E)}_F(r) = -\frac{\kappa^2q_e^2}{\Sigma^4(r)\left({G^2}_2 - {G^0}_0\right)}.
\end{equation}

\subsubsection{Static magnetic configurations}

For a static magnetic field, the nonzero components of the electromagnetic field tensor are $F^{23} = -F^{32}$. Based on the symmetry of the metric Eq.~\eqref{eq:general-metric}, the electromagnetic field tensor can be parameterized as
\begin{equation}
    F_{\mu\nu} = 2\delta^\theta_{[\mu}\delta^\varphi_{\nu]}b(r)\sin\theta.
\end{equation}
From the Bianchi identity ($\mathbf{dF} = 0$), we have
\begin{equation}
    0 = \mathbf{dF} = \frac{\dif}{\dif r}b(r)\sin\theta\ \dif r \wedge \dif\theta \wedge \dif\varphi.
\end{equation}
This demands that $b(r)$ must be a constant, which we identify as the magnetic charge $q_m$. The corresponding stress-energy tensor for the static magnetic field is therefore
\begin{equation}\label{eq:TSM}
    {{T^{(M)}}^\mu}_\nu = \text{diag}\left(L^{(M)}(r), L^{(M)}(r), L^{(M)}(r) - \frac{q_m^2 L^{(M)}_F(r)}{\Sigma^4(r)}, L^{(M)}(r) - \frac{q_m^2 L^{(M)}_F(r)}{\Sigma^4(r)}\right).
\end{equation}
Similarly, solving the field equations determines the underlying background scalar and NED functions. The specific form for $L^{(M)}_F(r)$ evaluates to
\begin{equation}\label{eq:LFM}
    L^{(M)}_F(r) = -\frac{\Sigma^4(r)\left({G^2}_2 - {G^0}_0\right)}{\kappa^2 q_m^2}.
\end{equation}
Comparing Eqs.~\eqref{eq:LFE} and~\eqref{eq:LFM} reveals a direct duality relationship between the electric and magnetic NED interpretations for a given background spacetime
\begin{equation}\label{eq:relationship-SE-SM}
    q^2_m L^{(M)}_F(r) = \frac{q^2_e}{L^{(E)}_F(r)} = -\frac{\Sigma^4(r)\left({G^2}_2 - {G^0}_0\right)}{\kappa^2}.
\end{equation}

\section{Perturbation equations in black-bounce spacetime}
\label{sec:Derivation}

To analyze axial gravitational perturbations, we consider linear metric fluctuations that impart a dragging of the inertial frame. The perturbed spacetime metric takes the form
\begin{equation}
\begin{split}
\label{eq:perturbed-metric}
    \dif s^2 = 
     f(r)\me^{2\delta\mu_0}(\dif t)^2 
     &- f(r)^{-1}\me^{2\delta\mu_1}(\dif r)^2 - \Sigma^2(r)\me^{2\delta\mu_2}(\dif \theta)^2\\
    & - \Sigma^2(r)\sin^2\theta \me^{2\delta\mu_3}(\dif \varphi - q_0 \dif t - q_1 \dif r - q_2 \dif \theta)^2,
\end{split}    
\end{equation}
where $\delta\mu_i$ and $q_i$ denote the spacetime perturbations. Setting these perturbations to zero cleanly recovers the static background metric, Eq.~\eqref{eq:general-metric}. The functions $q_i(t, r, \theta)$ specifically describe the odd-parity degrees of freedom, effectively imparting a slight rotation to the spacetime; hence, they correspond to the axial gravitational perturbations.

The governing master equations are derived by expanding the perturbed Einstein field equations to first order
\begin{equation}\label{eq:perturb-Einstein}
    \delta G_{\mu\nu} = \kappa^2 \delta T_{\mu\nu}.
\end{equation}
Here, $\delta G_{\mu\nu}$ represents the first-order variation of the Einstein tensor, which is fully parameterized by the perturbation variables in Eq.~\eqref{eq:perturbed-metric}. Concurrently, $\delta T_{\mu\nu}$ denotes the corresponding variation in the energy-momentum tensor of the supporting matter field. Because $\delta T_{\mu\nu}$ is structurally dependent on the nature of the background matter, the resulting perturbation equations will strictly vary across different field interpretations. 

Given the static and spherically symmetric nature of the background spacetime, there is no preferred axis of rotation. Consequently, any non-axisymmetric perturbation mode featuring an $\me^{im\varphi}$ dependence on the azimuthal angle $\varphi$ can be generated from the $m=0$ mode via a simple coordinate rotation~\cite{chandrasekhar1998mathematical}. Thus, without loss of generality, we set $m=0$ in the subsequent calculations.

To facilitate the separation of variables, it is highly advantageous to project the tensor equations onto an orthonormal tetrad frame
\begin{equation}
    \dif s^2 = \eta^{(a)(b)}e_{(a)\mu}e_{(b)\nu}\dif x^\mu \dif x^\nu,
\end{equation}
where 
\begin{equation}
    \eta^{(a)(b)} = \eta_{(a)(b)} = \text{diag}(1,-1,-1,-1),
\end{equation}
and the basis vectors are defined as
\begin{equation}
    \begin{split}
    &e_{(0)\mu} = (f(r)^{1/2}\me^{\delta\mu_0},0,0,0), \qquad 
    e_{(1)\mu} = (0,-f(r)^{-1/2}\me^{\delta\mu_1},0,0),\\
    &e_{(2)\mu} = (0,0,-\Sigma(r)\me^{\delta\mu_2},0), \\
    &e_{(3)\mu} = (q_0 \Sigma(r)\sin\theta \me^{\delta\mu_3}, q_1\Sigma(r)\sin\theta \me^{\delta\mu_3}, q_2 \Sigma(r)\sin\theta \me^{\delta\mu_3}, -\Sigma(r)\sin\theta \me^{\delta\mu_3}).
\end{split}
\end{equation}
Any tensor can then be transformed into this tetrad frame by contracting with the basis $e_{(a)\mu}$. For instance, the Einstein and energy-momentum tensors become
\begin{equation}
    G_{(a)(b)} = G^{\mu\nu}e_{(a)\mu}e_{(b)\nu}, \quad T_{(a)(b)} = T^{\mu\nu}e_{(a)\mu}e_{(b)\nu}.
\end{equation}

\subsection{Master equations for anisotropic fluid sources}

Varying the energy-momentum tensor of the anisotropic fluid defined in Eq.~\eqref{eq:TAF} yields the first-order perturbation
\begin{equation}\label{eq:delta-TAF}
\begin{split}
\delta T^{\mathrm{AF}}_{(a)(b)} =& (\rho+p_2)\delta(u_{(a)} u_{(b)}) + (\delta\rho+\delta p_2)(u_{(a)} u_{(b)})\\
&+ (p_1-p_2)\delta(x_{(a)} x_{(b)}) + (\delta p_1-\delta p_2)(x_{(a)} x_{(b)}) - \delta p_2 \eta_{(a)(b)}.
\end{split}
\end{equation}
For axial gravitational perturbations, the relevant off-diagonal tetrad components are $\delta T^{\mathrm{AF}}_{(1)(3)}$ and $\delta T^{\mathrm{AF}}_{(2)(3)}$. Enforcing the kinematic constraints on $x_\mu$ and $u_\mu$ outlined in Sec.~\ref{subsec:Afi}, we find that these matter perturbations vanish identically 
\begin{equation} 
    \delta T^{\mathrm{AF}}_{(1)(3)} = \delta T^{\mathrm{AF}}_{(2)(3)} = 0. 
\end{equation}
Expanding the corresponding components of the perturbed Einstein equations \eqref{eq:perturb-Einstein} in the tetrad frame and retaining only first-order terms, we obtain two coupled equations governing the axial degrees of freedom
\begin{equation}\label{eq:G13-AF}
    \left[\Sigma^2(r)f(r)\sin^3\theta\left(q_{1,\theta}-q_{2,r}\right)\right]_{,\theta} = -\Sigma^4(r)\sin^3\theta\left(q_{0,r}-q_{1,t}\right)_{,t},
\end{equation}
and
\begin{equation}\label{eq:G23-AF}
    f(r)\left[\Sigma^2(r)f(r)\sin^3\theta\left(q_{1,\theta}-q_{2,r} \right)\right]_{,r} = \Sigma^2(r)\sin^3\theta(q_{0,\theta}-q_{2,t})_{,t}.
\end{equation}
By cross-differentiating and separating variables, Eqs.~\eqref{eq:G13-AF} and~\eqref{eq:G23-AF} can be consolidated into a single master equation describing the axial gravitational perturbation for the anisotropic fluid interpretation
\begin{equation}
\label{eq:general-peq-afe1}
        \frac{\partial^2}{\partial r_*^2}H_2 - \frac{\partial^2}{\partial t^2}H_2 = \frac{f(r)}{\Sigma^2(r)}\left[(l-1)(l+2) - \Sigma^3(r)\frac{\partial}{\partial r}\left(\frac{f(r)}{\Sigma^2(r)}\frac{\partial}{\partial r}\Sigma(r)\right)\right]H_2,
\end{equation}
where the tortoise coordinate $r_*$ is defined by $\dif r_* \equiv \dif r/f(r)$, and the wave function $H_2$ is defined as
\begin{equation}\label{eq:H2}
    H_2 \equiv \frac{\sin^3\theta}{C^{-3/2}_{l+2}(\theta)}\Sigma(r)f(r)(q_{1,\theta}-q_{2,r}).
\end{equation}
Here, $C^{\nu}_{n}(\theta)$ is the Gegenbauer function with azimuthal quantum number $l$. This standard Schr\"odinger-like master equation is fully consistent with the forms derived in Refs.~\cite{Chen:2019iuo, Lessa:2024erf}.

\subsection{Master equations for nonlinear electrodynamics with scalar fields}

When the background spacetime is sourced by nonlinear electrodynamics coupled to a scalar field, the axial metric perturbations inevitably couple to the electromagnetic field perturbations. 

For static electric configurations, substituting Eqs.~\eqref{eq:TEM} and~\eqref{eq:Tphi} into the perturbed Einstein equations Eq.~\eqref{eq:perturb-Einstein} yields the relevant axial components:
\begin{equation}\label{eq:se-perturb-Einstein}
    \delta G_{(1)(3)} = \kappa^2 \delta T^{\mathrm{EM}}_{(1)(3)} = -\frac{\kappa^2 q_e}{\Sigma^2(r)}\delta F_{(0)(3)}, \quad \delta G_{(2)(3)} = 0.
\end{equation}
Notably, the scalar field perturbation strictly decouples from the odd-parity axial sector, exerting no influence on these equations. Expanding the first relation in Eq.~\eqref{eq:se-perturb-Einstein} within the tetrad frame provides the explicit form 
\begin{equation}\label{eq:G13-SE} 
\left[\Sigma^2(r)f(r)\sin^3\theta\left(q_{1,\theta}-q_{2,r}\right)\right]_{,\theta} + \Sigma^4(r)\sin^3\theta\left(q_{0,r}-q_{1,t}\right)_{,t} = 2\kappa^2 q_e\Sigma(r)\sqrt{f(r)}\sin^2\theta\delta F_{(0)(3)}.
\end{equation} 
The second relation in Eq.~\eqref{eq:se-perturb-Einstein} is structurally identical to Eq.~\eqref{eq:G23-AF}. However, the presence of the electromagnetic perturbation term $\delta F_{(0)(3)}$ introduces a new degree of freedom. This term is governed by the linearized nonlinear Maxwell equations \eqref{eq:maxwell-eq} and the Bianchi identity $\nabla_{[\rho}F_{\mu\nu]}=0$, yielding the required constraint
\begin{equation}\label{eq:F03-SE}
\begin{split}
    &\left[f(r)L^{(E)}_F(r)\left(\Sigma(r)\sqrt{f(r)}\delta F_{(0)(3)}\right)_{,r}\right]_{,r} + \frac{L^{(E)}_F(r)\sqrt{f(r)}}{\Sigma(r)}\left[\frac{\left(\sin\theta\delta F_{(0)(3)}\right)_{,\theta}}{\sin\theta}\right]_{,\theta}\\
    &-\frac{\Sigma(r)L^{(E)}_F(r)}{\sqrt{f(r)}}\delta F_{(0)(3),t,t} = q_e\sin\theta\left(q_{0,r}-q_{1,t}\right)_{,t}.
\end{split}
\end{equation}
By decoupling the angular dependence and separating variables in Eqs.~\eqref{eq:G13-SE} and~\eqref{eq:F03-SE}, we derive a system of two coupled master equations determining the simultaneous evolution of the axial gravitational and electric field perturbations
\begin{equation}\label{eq:SE-master-eq1}
     \frac{\partial^2}{\partial r_*^2}H^{(E)}_1 - \frac{\partial^2}{\partial t^2}H^{(E)}_1
     = V^{(E)}_{11}H^{(E)}_1 + V^{(E)}_{12}H_2,\\
\end{equation}
and
\begin{equation}\label{eq:SE-master-eq2}
     \frac{\partial^2}{\partial r_*^2}H_2 - \frac{\partial^2}{\partial t^2}H_2
     = V^{(E)}_{21}H^{(E)}_1 + V^{(E)}_{22}H_2,
\end{equation}
where the effective potential matrix elements are given by
\begin{subequations}
    \begin{equation}
        V^{(E)}_{11} = \frac{f(r)}{\Sigma^2(r)}\left[l(l+1) + \frac{\Sigma^2(r)}{2\sqrt{L^{(E)}_F(r)}}\frac{\dif}{\dif r}\left(\frac{f(r)}{\sqrt{L^{(E)}_F(r)}}\frac{\dif}{\dif r}L^{(E)}_F(r)\right) + \frac{16\pi q_e^2}{\Sigma^2(r)L^{(E)}_F(r)}\right],
    \end{equation}
    \begin{equation}
        V^{(E)}_{12} = \frac{2\sqrt{(l-1)(l+2)} q_e f(r)}{\Sigma^3(r)\sqrt{L^{(E)}_F(r)} },
    \end{equation}
    \begin{equation}
        V^{(E)}_{21} = \frac{\kappa^2\sqrt{(l-1)(l+2)} q_e f(r)}{\Sigma^3(r)\sqrt{L^{(E)}_F(r)}},
    \end{equation}
    \begin{equation}
        V^{(E)}_{22} = \frac{f(r)}{\Sigma^2(r)}\left[(l-1)(l+2) - \Sigma^3(r)\frac{\dif}{\dif r}\left(\frac{f(r)}{\Sigma^2(r)}\frac{\dif}{\dif r}\Sigma(r)\right)\right].
    \end{equation}
\end{subequations}
Here, the electromagnetic wave function is defined as
\begin{equation}\label{eq:SE-H1}
    H^{(E)}_1 = -\frac{2\sqrt{(l-1)(l+2)}\sin\theta}{3C^{-1/2}_{l+1}(\theta)}\Sigma(r)\sqrt{f(r)L^{(E)}_F(r)}\delta F_{(0)(3)},
\end{equation}
and $H_2$ remains defined as in Eq.~\eqref{eq:H2}. Due to the complexity and nonzero off-diagonal components $V^{(E)}_{12}$ and $V^{(E)}_{21}$, Eqs.~\eqref{eq:SE-master-eq1} and~\eqref{eq:SE-master-eq2} represent an irreducibly coupled two-channel system.

For static magnetic configurations, following the same perturbative procedure yields the relevant field equations
\begin{equation}\label{eq:sm-perturb-Einstein}
    \delta G_{(1)(3)} = \kappa^2 \delta T^{\mathrm{EM}}_{(1)(3)} = -\frac{\kappa^2 q_m L^{(M)}_F(r)}{\Sigma^2(r)}\delta F_{(1)(2)}, \quad \delta G_{(2)(3)} = 0.
\end{equation}
The explicit form of the first equation expands to
\begin{equation}\label{eq:G13-SM}
\begin{split}   \left[\Sigma^2(r)f(r)\sin^3\theta\left(q_{1,\theta}-q_{2,r}\right)\right]_{,\theta} 
&+ \Sigma^4(r)\sin^3\theta\left(q_{0,r}-q_{1,t}\right)_{,t} \\
&= 2\kappa^2q_m\Sigma(r)L^{(M)}_F(r)\sqrt{f(r)}\sin^2\theta\delta F_{(1)(2)}.
\end{split} 
\end{equation}
The constraint condition for $\delta F_{(1)(2)}$ derived from the nonlinear Maxwell equations reads
\begin{equation}
\begin{split}\label{eq:F12-SM}
    &\left[\frac{f(r)}{L^{(M)}_F(r)}\left(\Sigma(r)\sqrt{f(r)}L^{(M)}_F(r)\delta F_{(1)(2)}\right)_{,r}\right]_{,r} + \frac{\sqrt{f(r)}}{\Sigma(r)} \left[\frac{\left(\sin\theta\delta F_{(1)(2)}\right)_{,\theta}}{\sin\theta}\right]_{,\theta}\\
    &-\frac{\Sigma(r)}{\sqrt{f(r)}} \delta F_{(1)(2),t,t} = -q_m\sin\theta\left(q_{0,r}-q_{1,t}\right)_{,t}.
\end{split}
\end{equation}
Separating variables in Eqs.~\eqref{eq:G13-SM} and~\eqref{eq:F12-SM} produces the coupled master equations for the magnetic configuration
\begin{equation}\label{eq:SM-master-eq1}
     \frac{\partial^2}{\partial r_*^2}H^{(M)}_1 - \frac{\partial^2}{\partial t^2}H^{(M)}_1
     = V^{(M)}_{11}H^{(M)}_1 + V^{(M)}_{12}H_2,\\
\end{equation}
and
\begin{equation}\label{eq:SM-master-eq2}
     \frac{\partial^2}{\partial r_*^2}H_2 - \frac{\partial^2}{\partial t^2}H_2
     = V^{(M)}_{21}H^{(M)}_1 + V^{(M)}_{22}H_2,
\end{equation}
where the corresponding potential matrix elements evaluate to
\begin{subequations}
    \begin{equation}
    \begin{split}
        V^{(M)}_{11} =& \frac{f(r)}{\Sigma^2(r)}\left\{l(l+1) + \frac{\Sigma^2(r)\sqrt{L^{(M)}_F(r)}}{2}\frac{\dif}{\dif r}\left[f(r)\sqrt{L^{(M)}_F(r)}\frac{\dif}{\dif r}\left(\frac{1}{L^{(M)}_F(r)}\right)\right]\right.\\
        &\left.+ 16\pi q_m^2\frac{ L^{(M)}_F(r)}{\Sigma^2(r)}\right\},
    \end{split}
    \end{equation}
    \begin{equation}
        V^{(M)}_{12} = \frac{2\sqrt{(l-1)(l+2)} q_m f(r)\sqrt{L^{(M)}_F(r)}}{\Sigma^3(r)},
    \end{equation}
    \begin{equation}
        V^{(M)}_{21} = \frac{\kappa^2\sqrt{(l-1)(l+2)} q_m f(r)\sqrt{L^{(M)}_F(r)}}{\Sigma^3(r)},
    \end{equation}
    \begin{equation}
        V^{(M)}_{22} = \frac{f(r)}{\Sigma^2(r)}\left[(l-1)(l+2) - \Sigma^3(r)\frac{\dif}{\dif r}\left(\frac{f(r)}{\Sigma^2(r)}\frac{\dif}{\dif r}\Sigma(r)\right)\right].
    \end{equation}
\end{subequations}
The magnetic wave function is defined as
\begin{equation}\label{eq:SM-H1}
    H^{(M)}_1 = -\frac{2\sqrt{(l-1)(l+2)}\sin\theta}{3C^{-1/2}_{l+1}(\theta)}\Sigma(r)\sqrt{f(r)L^{(M)}_F(r)}\delta F_{(1)(2)},
\end{equation}
with $H_2$ remaining unmodified.

Comparing the master equations for the static electric and magnetic configurations reveals a striking duality. The most immediate observation is that the purely gravitational effective potential $V_{22}$ is completely identical across both frameworks 
\begin{equation} V^{(E)}_{22} = \frac{f(r)}{\Sigma^2(r)}\left[(l-1)(l+2) - \Sigma^3(r)\frac{\dif}{\dif r}\left(\frac{f(r)}{\Sigma^2(r)}\frac{\dif}{\dif r}\Sigma(r)\right)\right] = V^{(M)}_{22} \equiv V_{22}. 
\end{equation}
Furthermore, leveraging the background duality relationship linking the electric and magnetic interpretations Eq.~\eqref{eq:relationship-SE-SM}, we discover that the remaining elements of the potential matrix are also functionally identical when expressed purely in terms of the geometric background invariants ${G^0}_0$ and ${G^2}_2$
 \begin{equation}
    \begin{split}
        V^{(E)}_{11} =& \frac{f(r)}{\Sigma^2(r)}\left\{l(l+1) + \frac{\Sigma^4(r)\sqrt{{G^0}_0-{G^2}_2}}{2}\frac{\dif}{\dif r}\left\{f(r)\Sigma^2(r)\sqrt{{G^0}_0-{G^2}_2}\frac{\dif}{\dif r}\left[\frac{1}{\Sigma^4(r)\left({G^0}_0-{G^2}_2\right)}\right]\right\}\right.\\
        &\left.+ 2\Sigma^2(r)\left({G^0}_0-{G^2}_2\right)\right\}\\
        &= V^{(M)}_{11} \equiv V_{11},
    \end{split}
\end{equation}
\begin{equation}
        V^{(E)}_{12} = \frac{2\sqrt{(l-1)(l+2)} f(r)\sqrt{{G^0}_0-{G^2}_2}}{\kappa\Sigma(r)} = V^{(M)}_{12} \equiv V_{12},
\end{equation}
\begin{equation}
\label{eq:Vcomp}        V^{(E)}_{21} = \frac{\kappa\sqrt{(l-1)(l+2)} f(r)\sqrt{{G^0}_0-{G^2}_2}}{\Sigma(r)} = V^{(M)}_{21} \equiv V_{21}.
\end{equation}
We denote these universal effective potentials generally as $V_{ij}$ ($i,j=1,2$). From this rigorous equivalence, we draw two fundamental conclusions regarding the axial perturbations: 
\begin{itemize}
    \item The effective potential matrix dictating the evolution of axial gravitational perturbations is uniquely determined by the geometric structure of the metric itself. It is entirely insensitive to whether the underlying nonlinear electrodynamic field is sourced by electric or magnetic charges.
    \item Consequently, the coupled dynamical evolution of axial and electromagnetic perturbations is universally identical for both static electric and static magnetic configurations.
\end{itemize}
Therefore, in our subsequent analysis, evaluating the influence of the NED field on axial gravitational perturbations for either the electric or magnetic case automatically guarantees applicability to the other.

\section{Initial and boundary conditions for perturbations}
\label{sec:InBound}

As established in Sec.~\ref{sec:Derivation}, different matter-field interpretations of the black-bounce spacetime give rise to structurally distinct master equations governing axial gravitational perturbations. In this section, we investigate the time-domain evolution of these equations by numerically evolving the perturbation wave functions $H_1$ and $H_2$ under physically well-motivated initial and boundary conditions.

\subsection{Initial perturbation setup}

The numerical analysis is carried out within a physical setup in which a nonvanishing axial gravitational perturbation is present at the initial time $t = 0$, after which the perturbation evolves freely on the fixed black-bounce background in the absence of external sources. This configuration is standard in time-domain studies of black hole and wormhole perturbations, as it isolates the intrinsic dynamical response of the spacetime geometry from any source-dependent transients.

At $t = 0$, the perturbation function $H_1$ is initialized as a localized Gaussian wave packet,
\begin{equation}
H_1(t=0,\, r_*) = A \exp\!\left[-\frac{(r_* - r_{*0})^2}{\sigma^2}\right],
\label{eq:IC_H1}
\end{equation}
where $A$ is the amplitude, $r_{*0}$ specifies the initial centroid of the wave packet in the tortoise coordinate $r_*$, and $\sigma$ controls its spatial width. The complementary perturbation function is set to zero initially,
\begin{equation}
H_2(t=0,\, r_*) = 0,
\label{eq:IC_H2}
\end{equation}
corresponding to the absence of any initial excitation in the gravitational channel. This choice is particularly well suited to probing the response of the coupled NED system, since any nontrivial late-time evolution in $H_2$ must arise entirely through dynamical coupling to $H_1$ via the off-diagonal potentials $V_{12}$ and $V_{21}$. The initial time derivatives of both perturbation functions are then determined self-consistently from the master equations,  Eqs.~\eqref{eq:IC_H1}--\eqref{eq:IC_H2}.

\subsection{Boundary conditions and quasinormal mode extraction}

The master equations derived in Sec.~\ref{sec:Derivation} are solved subject to boundary conditions imposed at the two asymptotic ends of the tortoise coordinate $r_*$. At spatial infinity, $r_* \to +\infty$, purely outgoing wave conditions are enforced, ensuring that no incoming radiation enters the system from the exterior. At the opposite boundary, $r_* \to -\infty$, purely ingoing wave conditions are imposed.

The physical interpretation of the $r_* \to -\infty$ boundary depends on the causal structure of the spacetime. In the black-hole branch ($a \leq 2M$), this limit corresponds to the event horizon, where regularity requires only ingoing modes. In the wormhole branch ($a > 2M$), the same limit instead represents the asymptotically flat region on the far side of the throat, and the ingoing-wave condition is adopted consistently in both cases as the physically natural choice.

Under these boundary conditions, the characteristic complex oscillation frequencies that emerge from the time evolution correspond to the QNMs, which encode the dissipative dynamical response of the black-bounce spacetime to axial gravitational perturbations. Specifically, the real and imaginary parts of a QNM frequency $\omega = \omega_R + \mi\,\omega_I$ determine, respectively, the oscillation frequency and the decay rate (with $\omega_I < 0$ indicating mode stability).

\subsection{Numerical time evolution and source dependence}
\label{sec:numerics}

The time evolution of the perturbation equations is performed using a finite-difference scheme in the time domain~\cite{Konoplya:2011qq}. This approach enables a direct numerical simulation of the propagation and decay of the perturbation wave packets and yields time-series data for the wave functions at a fixed extraction point $r_*^{\rm obs}$.

The two matter-source interpretations considered in this work lead to qualitatively different dynamical structures. Under the AF interpretation, the axial gravitational perturbation is governed by the single Schr\"odinger-like master equation Eq.~\eqref{eq:general-peq-afe1} for $H_2$ alone, so the time evolution involves only one independent degree of freedom.

In contrast, when the black-bounce spacetime is supported by NED coupled to a scalar field, the axial perturbations are described by the coupled system of master equations~\eqref{eq:SE-master-eq1}--\eqref{eq:SE-master-eq2} (or equivalently~\eqref{eq:SM-master-eq1}--\eqref{eq:SM-master-eq2} for the magnetic configuration). In this case, $H_1$ and $H_2$ are dynamically coupled throughout the entire evolution via the off-diagonal potentials $V_{12}$ and $V_{21}$, and their interplay plays a decisive role in shaping the late-time ringdown behavior, as analyzed in detail in Sec.~\ref{sec:Dynamical}.

The QNM frequencies are extracted from the resulting numerical waveforms using the Prony method~\cite{Berti:2009kk}, which is well suited for decomposing exponentially damped oscillatory signals during the ringdown phase into a discrete sum of complex exponentials. This procedure yields accurate determinations of the complex quasinormal frequencies from the time-domain data, and all extracted values are tabulated in the supplementary material for verification.

\section{Dynamical evolution of axial perturbations}
\label{sec:Dynamical}

In this section, we analyze the axial gravitational perturbations in the SV black-bounce spacetime, with the primary aim of clarifying how different matter-field interpretations of the same background geometry influence the dynamical evolution and QNM spectra. The numerical setup follows the procedure outlined in Sec.~\ref{sec:InBound}, and the results are organized according to the two geometric branches defined by the bounce parameter $a$.

\subsection{Simpson--Visser metric and BH/WH branches}

The SV black-bounce spacetime is characterized by the metric functions given in Eq.~\eqref{eq:par_sv}, which smoothly interpolate between a regular black hole and a traversable wormhole as $a$ increases. Specifically, for $a \leq 2M$ the spacetime possesses an event horizon and describes a regular black hole, whereas for $a > 2M$ the horizon is absent and the geometry transitions into a traversable wormhole with a throat at $r = 0$.

Although the background metric is uniquely specified by Eq.~\eqref{eq:par_sv}, the underlying matter content is not. As discussed in Sec.~\ref{sec:Spacetime}, the SV geometry can be supported either by an AF or by NED coupled to a scalar field, in both electric and magnetic configurations. As we demonstrate below, this non-uniqueness of the matter source has direct and observable consequences for the dynamics of perturbations.

\subsubsection{Effective potentials for axial gravitational perturbations}

The two matter-source interpretations give rise to structurally distinct effective potentials, reflecting the different perturbation equations derived in Sec.~\ref{sec:Derivation}.

Under the AF interpretation, the axial gravitational perturbation is governed by a single Schr\"odinger-like master equation for $H_2$, with the effective potential
\begin{equation}
\begin{split}
    V^{(\mathrm{AF})}
    =\,&\frac{f(r)}{\Sigma^2(r)}\left[\mu^2-\Sigma^3(r)\frac{\mathrm{d}}{\mathrm{d} r}\left(\frac{f(r)}{\Sigma^2(r)}\frac{\mathrm{d}}{\mathrm{d} r}\Sigma(r)\right)\right]\\
    =\,&\frac{\left(\sqrt{r^2+a^2}-2 M\right) \left\{a^2 \left[\left(\mu ^2-1\right) \sqrt{r^2+a^2}+2 M\right]+r^2 \left[\left(\mu ^2+2\right) \sqrt{r^2+a^2}-6 M\right]\right\}}{\left(r^2+a^2\right)^3},
\end{split}
\end{equation}
where $\mu^2 \equiv (l+2)(l-1)$ and the tortoise coordinate $r_*$ is defined via $\mathrm{d}r_* = \mathrm{d}r/f(r)$. This single-channel potential is the direct analog of the Regge--Wheeler potential for a Schwarzschild black hole, recovered in the limit $a \to 0$.

Under the NED interpretation, the axial sector involves a genuinely coupled two-channel system for $(H_1, H_2)$, as derived in Sec.~\ref{sec:Derivation}. The corresponding $2\times 2$ effective potential matrix has elements
\begin{subequations}
\label{eq:potential_matrix}
\begin{align}
\label{eq:V11}
    V_{11} =\,&\frac{f(r)}{\Sigma^4(r)}\left[(\mu^2+2)\Sigma^2(r)+\frac{\Sigma^4(r)}{2\sqrt{L^{(e)}_F(r)}}\frac{\mathrm{d}}{\mathrm{d} r}\!\left(\frac{f(r)}{\sqrt{L^{(e)}_F(r)}}\frac{\mathrm{d}}{\mathrm{d} r}L^{(e)}_F(r)\right)+\frac{16\pi q_e^2}{L^{(e)}_F(r)}\right]\notag\\
    =\,&\frac{\left(\sqrt{a^2+r^2}-2 M\right)}{4 \left(a^2+r^2\right)^{7/2}}\Big[2 a^4 \left(2 \mu ^2+5\right)+a^2 \left(20 M \sqrt{a^2+r^2}+\left(8 \mu ^2+17\right) r^2\right)\notag\\
    &\hspace{6.8cm}+6 M r^2 \sqrt{a^2+r^2}+\left(4 \mu ^2+7\right) r^4\Big],\\[6pt]
    V_{12} =\,&\frac{2\mu q_e f(r)}{\Sigma^3(r)\sqrt{L^{(e)}_F(r)}}
        =\frac{\sqrt{\dfrac{3}{2 \pi }}\, \mu  \sqrt{a^2 M}\, \left(\sqrt{a^2+r^2}-2 M\right)}{\left(a^2+r^2\right)^{9/4}}=\frac{1}{4\pi}V_{21},\\[6pt]
    V_{22} =\,&\frac{f(r)}{\Sigma^2(r)}\left[\mu^2-\Sigma^3(r)\frac{\mathrm{d}}{\mathrm{d} r}\!\left(\frac{f(r)}{\Sigma^2(r)}\frac{\mathrm{d}}{\mathrm{d} r}\Sigma(r)\right)\right] = V^{(\mathrm{AF})}.
\end{align}
\end{subequations}
Several features of this potential matrix deserve emphasis. First, the diagonal element $V_{22}$ is \emph{identical} to the AF potential $V^{(\mathrm{AF})}$, confirming that the purely gravitational sector is insensitive to the matter-source interpretation at the level of the background geometry. Second, the off-diagonal elements $V_{12}$ and $V_{21}$ are nonzero whenever the charge $q_e \neq 0$ and $l \geq 2$, reflecting the irreducible coupling between the gravitational and electromagnetic perturbation channels. Third, and most importantly, the full potential matrix—including the EM-sector diagonal $V_{11}$—is entirely determined by the spacetime geometry via the background Einstein-tensor components, independently of whether the NED source is electric or magnetic, cf.\ Eq.~\eqref{eq:relationship-SE-SM}. The non-symmetric structure $V_{12} \neq V_{21}$ (differing by a factor of $4\pi$) is a direct manifestation of the non-Hermitian character of the coupled system, which will be analyzed in detail in Sec.~\ref{sec:NH}.

\subsubsection{Eikonal limit and photon-sphere correspondence}
\label{sec:eikonal}
In the limit of large multipole number $l\gg 1$, quasinormal frequencies admit an analytic approximation controlled by the properties of unstable circular null orbits (the photon sphere). For a general spherically symmetric metric of the form Eq.~\eqref{eq:general-metric}, the null geodesic effective potential is
\begin{equation}\label{eq:Vnull}
V_{\mathrm{eff}}^{\mathrm{null}}(r) = \frac{f(r)}{\Sigma^2(r)},
\end{equation}
and the photon-sphere radius $r_c$ satisfies $\left.\tfrac{d}{dr}V_{\mathrm{eff}}^{\mathrm{null}}\right|_{r=r_c}=0$, which for the Simpson--Visser metric Eq.~\eqref{eq:par_sv} yields the elegant condition
\begin{equation}\label{eq:photonsphere}
\Sigma(r_c) = \sqrt{r_c^2+a^2} = 3M, \qquad\text{i.e.}\quad r_c = \sqrt{9M^2 - a^2}\quad (a\leq 2M).
\end{equation}
The photon-sphere orbital frequency and Lyapunov exponent are
\begin{equation}\label{eq:OmegaLyap}
\Omega_c = \sqrt{\frac{f(r_c)}{\Sigma^2(r_c)}} = \frac{1}{3\sqrt{3}\,M}, \qquad \lambda_L = \frac{1}{9 M^2}\sqrt{\frac{9M^2-a^2}{3}},
\end{equation}
both of which are independent of $l$. In the eikonal approximation, the QNM frequencies are given by~\cite{Cardoso:2008bp}
\begin{equation}\label{eq:eikonal}
\omega_n^{\mathrm{(eik)}} = \Omega_c\!\left(l+\tfrac{1}{2}\right) - \mi\lambda_L\!\left(n+\tfrac{1}{2}\right) + \mathcal{O}(l^{-1}), \qquad n=0,1,2,\ldots
\end{equation}

Two immediate consequences arise from this formulation. First, Eq.~\eqref{eq:photonsphere} demonstrates that $r_c$, and consequently $\Omega_c$, depend strictly on the metric functions $f(r)$ and $\Sigma(r)$, completely independent of the matter-field interpretation. Therefore, the eikonal QNM spectrum is \emph{the same} for both the AF and NED sources. Isospectrality is asymptotically restored in the limit $l\to\infty$, implying that the matter-source discrimination reported in Sec.~\ref{subsec:qnm_tables} is purely a finite-$l$ effect that scales as $\mathcal{O}(l^{-2})$.

Second, the EM-sector potential $V_{11}$ contains an additional charge-dependent term Eq.~\eqref{eq:V11} beyond the universal $l(l+1)f/\Sigma^2$ factor. While the geometric photon sphere remains invariant, this additional term amplifies the effective potential barrier for the EM-led mode at finite $l$. This raises the effective photon-sphere frequency, yielding the ratio
\begin{equation}\label{eq:omRatio}
\frac{\omega_R^{(\mathrm{EM})}}{\omega_R^{(\mathrm{GR})}} \approx \sqrt{1 + \frac{16\pi q_e^2}{l(l+1)\Sigma^4(r_c)L_F^{(E)}(r_c)}}.
\end{equation}
For the parameters of the wormhole branch ($a>2.14M$, $l=2$), this ratio evaluates to approximately $1.4$. This provides a rigorous analytical origin for the empirical scaling relation observed in our numerical data for $a>2.14M$, where $\omega_{\rm EM} \approx 1.4\,\omega_{\rm AF}$.

Table~\ref{tab:eikonal} compares the eikonal prediction \eqref{eq:eikonal} with the numerically extracted AF fundamental ($n=0$) mode for several values of $a$. The fractional discrepancy $\varepsilon = |\omega_0^{(\mathrm{AF})}-\omega_0^{(\mathrm{eik})}| / |\omega_0^{(\mathrm{AF})}|$ is $\lesssim 3\%$ for the real part and $\lesssim 8\%$ for the imaginary part at $l=2$, which is highly consistent with the expected $\mathcal{O}(1/l^2)$ corrections. These results confirm that the eikonal formula provides a reliable qualitative guide to the QNM spectrum, while achieving quantitative precision still necessitates the full numerical treatment detailed in Sec.~\ref{sec:numerics}.

\begin{table}[!ht]
    \centering
 \begin{tabular}{cccccc}\hline
    $a/M$   &  $\omega_0^{(\mathrm{AF})}$  & $\omega_0^{(\mathrm{eik})}$                         & 
         $\varepsilon_R$ & $\varepsilon_I$                                         \\ \hline
    $0.2$    & 0.373094 - 0.0904845 i     & 0.3849 - 0.0960 i   &  3.2\%  &   6.1\%    \\ \hline
    $0.4$    & 0.373074 - 0.0897516 i     & 0.3849 - 0.0957 i   &  3.2\%  &   6.6\%    \\ \hline
    $0.6$    & 0.373013 - 0.0885294 i     & 0.3849 - 0.0952 i   &  3.2\%  &   7.6\%    \\ \hline
    $0.8$    & 0.372903 - 0.0868313 i     & 0.3849 - 0.0944 i   &  3.2\%  &   8.8\%    \\ \hline
    $1.0$    & 0.372748 - 0.0845463 i     & 0.3849 - 0.0933 i   &  3.3\%  &   10.4\%    \\ \hline
    $1.2$    & 0.372460 - 0.0816476 i     & 0.3849 - 0.0918 i   &  3.3\%  &   12.5\%    \\ \hline
    $1.4$    & 0.372216 - 0.0782923 i     & 0.3849 - 0.0900 i   &  3.4\%  &   14.9\%    \\ \hline
    $1.6$    & 0.371558 - 0.0741851 i     & 0.3849 - 0.0878 i   &  3.6\%  &   18.3\%    \\ \hline
    $1.8$    & 0.370589 - 0.0691939 i     & 0.3849 - 0.0852 i   &  3.8\%  &   23.1\%    \\ \hline
    $2.0$    & 0.369130 - 0.0641531 i     & 0.3849 - 0.0821 i   &  4.3\%  &   27.9\%    \\ \hline
  \end{tabular}
      \captionsetup{width=.9\textwidth}
\caption{Eikonal prediction \eqref{eq:eikonal} versus numerically extracted AF axial QNMs ($l=2$, $n=0$). The fractional error $\varepsilon_R$ ($\varepsilon_I$) is computed from the real (imaginary) parts.}
\label{tab:eikonal}
\end{table}

\subsection{Quasinormal mode spectra across matter interpretations}
\label{subsec:qnm_tables}

We systematically extracted the axial gravitational QNM frequencies from the time-domain signals across both the black-hole ($a \leq 2M$) and wormhole ($a > 2M$) branches. The comprehensive dataset is tabulated in the Supplementary Material. For each value of the bounce parameter $a$, we track four distinct mode families: (i) the QNM under the AF interpretation, governed by a single master equation for the axial gravitational variable $H_2$; (ii) the gravitationally led (GR-led) and (iii) electromagnetically led (EM-led) QNMs under the NED interpretation, which emerge from the irreducibly coupled two-channel system $(H_1, H_2)$; and (iv) the pure electromagnetic (EM) QNM, computed by artificially isolating the electromagnetic field from the gravitational perturbations. Across all scenarios, the imaginary parts of the frequencies remain strictly negative, confirming the linear mode stability of the black-bounce spacetime regardless of the underlying matter-source interpretation.

We note a specific phenomenological feature in the transitional regime $2.0 < a/M < 2.10$: here, the time-domain waveforms exhibit pronounced echo interference that obscures the primary exponential decay. Consequently, a uniquely defined fundamental ringdown frequency cannot be reliably extracted via standard Prony fitting. These specific parameter points are marked as ``unidentified'' in the Supplementary Material.

\begin{figure}[!ht]
    \centering
    \begin{subfigure}[b]{0.32\textwidth}
        \centering
        \includegraphics[width=\textwidth]{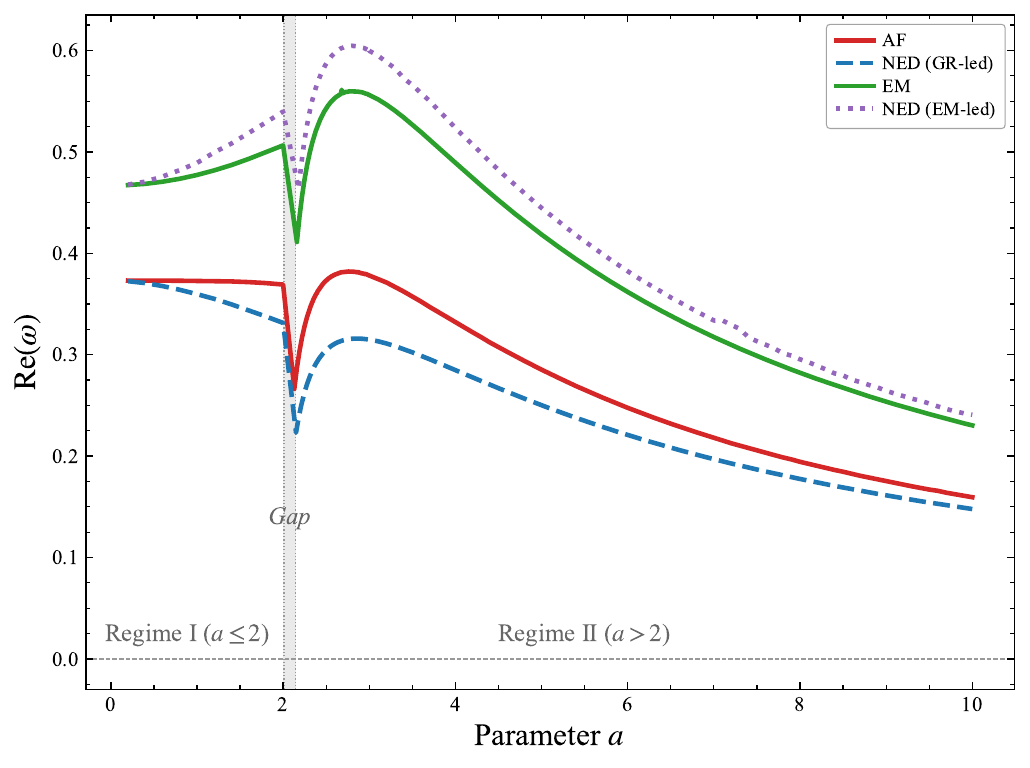}
        \caption{}
        \label{fig:qnms_a_real}
    \end{subfigure}
    \hfill
    \begin{subfigure}[b]{0.32\textwidth}
        \centering
        \includegraphics[width=\textwidth]{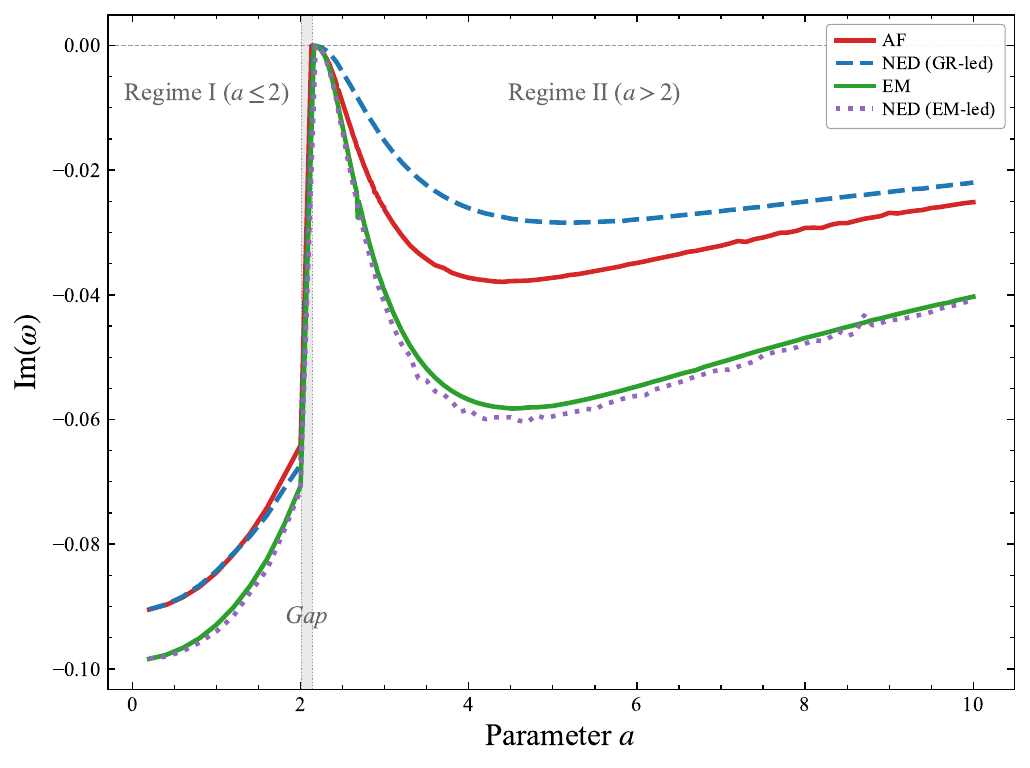}
        \caption{}
        \label{fig:qnms_a_imag}
    \end{subfigure}
    \hfill
    \begin{subfigure}[b]{0.32\textwidth}
        \centering
        \includegraphics[width=\textwidth]{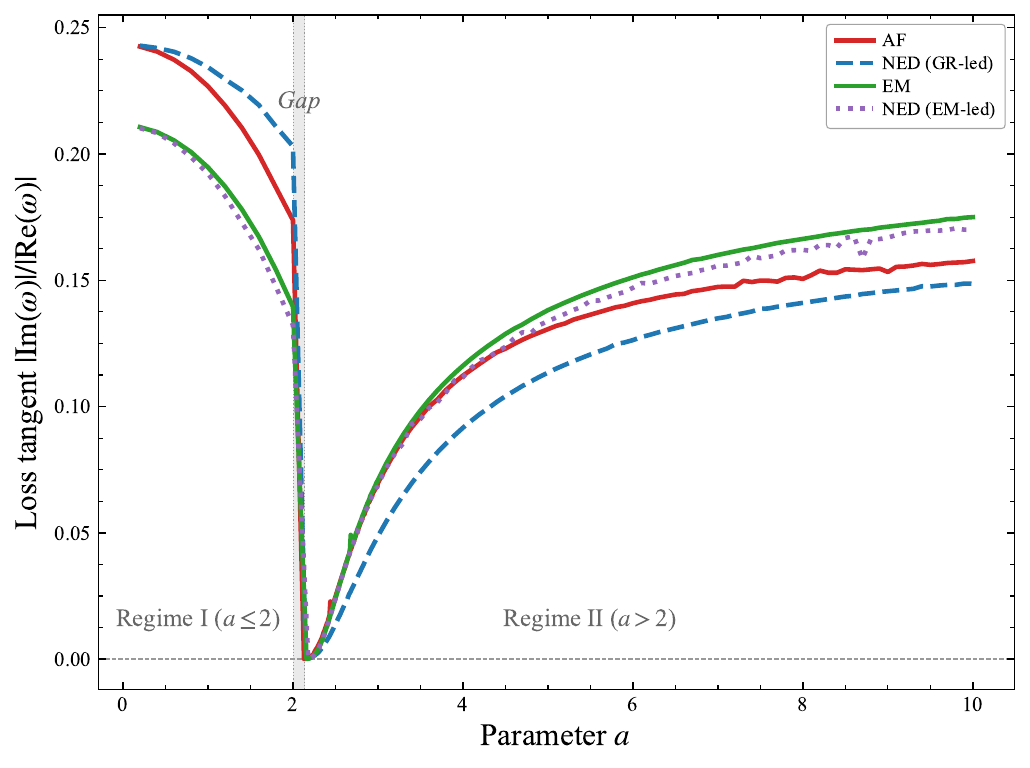}
        \caption{}
        \label{fig:qnms_a_loss}
    \end{subfigure}
    \captionsetup{width=.9\textwidth}
    \caption{QNM spectra as functions of the bounce parameter $a$ (in units of $M$, $l=2$) for the four mode families: AF (solid red), NED GR-led (dashed blue), EM (solid green), and NED EM-led (dotted purple). The vertical gray band indicates the geometric transition at $a=2M$. Panels \subref{fig:qnms_a_real}, \subref{fig:qnms_a_imag}, and \subref{fig:qnms_a_loss} display the real part, imaginary part, and loss tangent ($|\Im(\omega)|/\Re(\omega)$) of the fundamental frequencies, respectively.}
    \label{fig:qnms_a}
\end{figure}

\paragraph{Black-hole branch ($a \leq 2M$).}
The evolution of the QNM spectra as a function of $a$ is presented in Fig.~\ref{fig:qnms_a}, where the vertical gray band marks the geometric phase transition at $a=2M$. In the black-hole branch, the AF frequencies exhibit mild variation: the real part $\Re(\omega_{\rm AF})$ remains nearly constant, while the magnitude of the imaginary part $|\Im(\omega_{\rm AF})|$ decreases monotonically, reflecting a gradual reduction in the damping rate. The EM and NED (EM-led) families maintain markedly higher oscillation frequencies and larger decay rates throughout this regime. 

The NED (GR-led) trajectory tracks the AF value closely at small $a$, but a systematic deviation develops as $a$ approaches the $2M$ threshold. Both real parts drop sharply near the transition, but $\Re(\omega_{\rm NED})$ falls strictly below $\Re(\omega_{\rm AF})$. Simultaneously, the damping rate becomes slightly amplified
\begin{equation}
\label{eq:GRfaster}
|\Im(\omega_{\rm NED})| \gtrsim |\Im(\omega_{\rm AF})| \qquad (a \leq 2M).
\end{equation}
As depicted in the middle panel of Fig.~\ref{fig:qnms_a}, this implies that the GR-led NED ringdown decays \emph{faster} than in the single-channel AF case, a consequence of the enhanced leakage through the coupled electromagnetic channel. 

\paragraph{Wormhole branch ($a > 2M$).}
Crossing into the wormhole branch triggers a qualitatively distinct dynamic, signaling the onset of subradiant-like interference. While the EM-led modes remain comparatively high-frequency and strongly damped, the GR-led NED fundamental mode becomes simultaneously \emph{lower in frequency} and \emph{less damped} than its AF counterpart
\begin{equation}
\label{eq:WHslower}
\Re(\omega_{\rm NED}) < \Re(\omega_{\rm AF}),
\qquad
|\Im(\omega_{\rm NED})| < |\Im(\omega_{\rm AF})|
\qquad (a > 2M).
\end{equation}
This branch-dependent inversion is most clearly visualized in the middle panel of Fig.~\ref{fig:qnms_a}, where the ordering of the decay rates firmly flips. Hence, in the wormhole branch, the dominant late-time NED ringdown survives \emph{longer} than the pure-gravity AF signal—a direct reversal of the behavior observed in the black-hole regime. 

To provide a dimensionless measure of this relative damping, we evaluate the loss tangent $|\Im(\omega)|/\Re(\omega)$ (Fig.~\ref{fig:qnms_a}, right panel). This metric renders the crossover exceptionally transparent: the NED (GR-led) loss tangent exceeds the AF baseline in the BH branch but systematically falls below it throughout the WH branch. Meanwhile, the EM-led loss tangent remains the largest across the entire parameter space, firmly establishing its role as the highly radiative eigenmode of the coupled system.

\begin{figure}[!ht]
    \centering
    \begin{subfigure}[b]{0.49\textwidth}
        \centering
        \includegraphics[width=\textwidth]{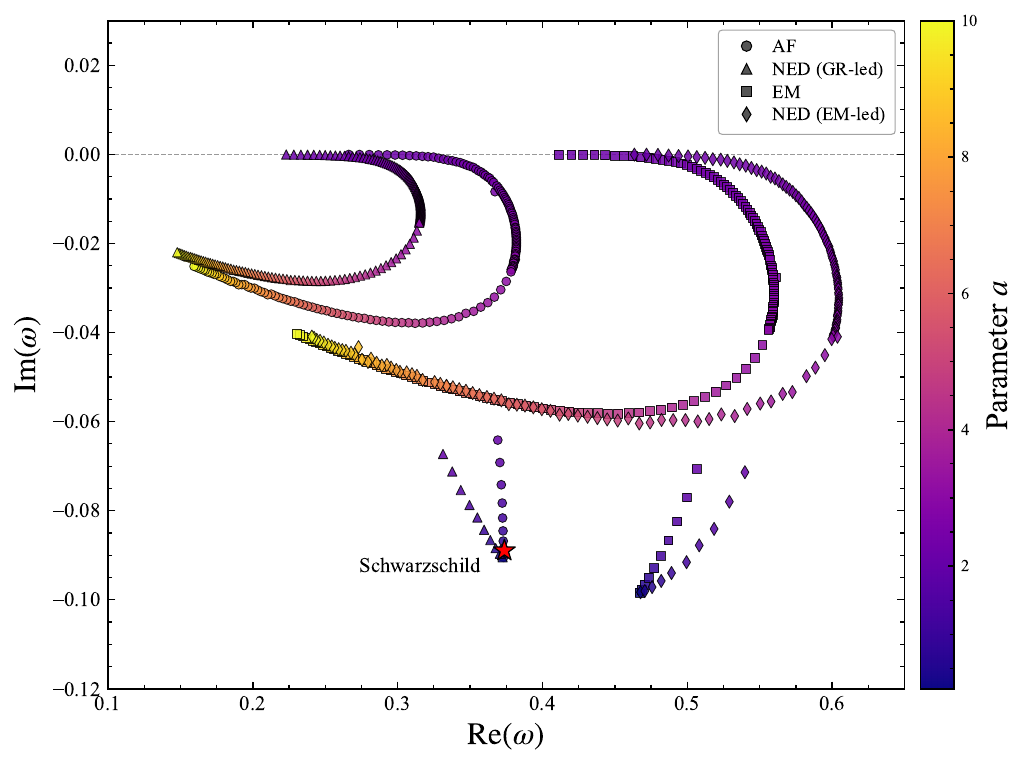}
        \caption{}
        \label{fig:qnms_complex_combined}
    \end{subfigure}
    \hfill
    \begin{subfigure}[b]{0.49\textwidth}
        \centering
        \includegraphics[width=\textwidth]{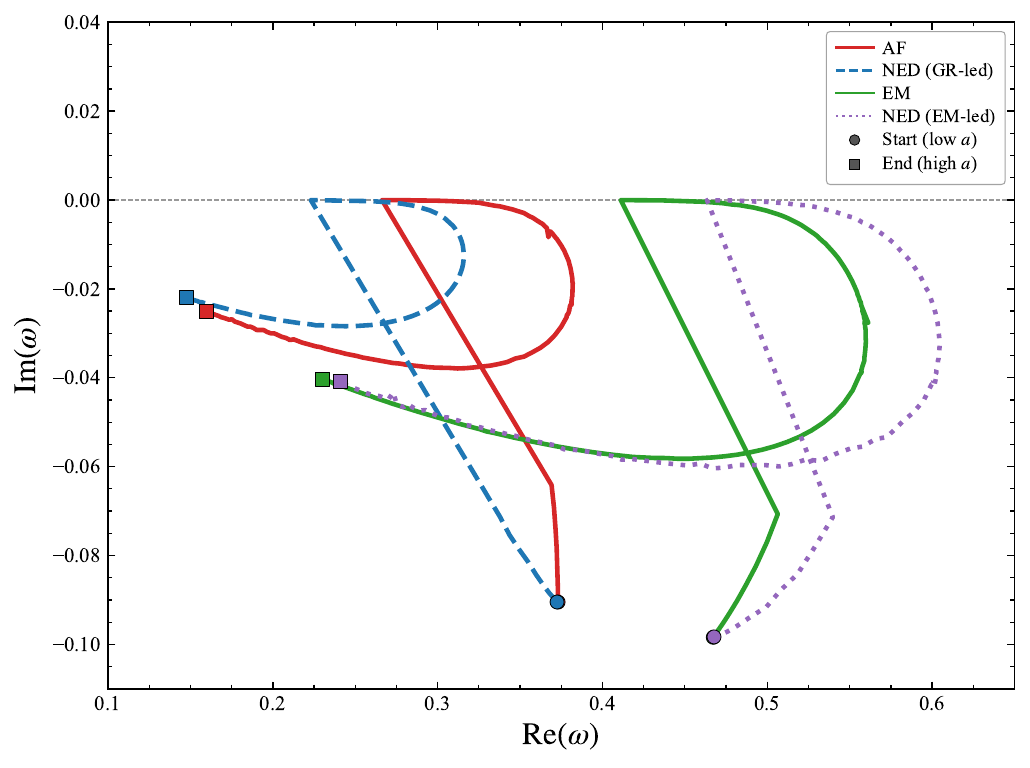}
        \caption{}
        \label{fig:qnms_complex_argand}
    \end{subfigure}
    \captionsetup{width=.9\textwidth}
    \caption{Trajectories of the QNM frequencies in the complex plane as the bounce parameter $a$ varies across the BH-WH transition ($l = 2$). Panel \subref{fig:qnms_complex_combined} shows the combined spectral trajectories for all four mode families. Panel \subref{fig:qnms_complex_argand} presents the corresponding Argand diagram, colored continuously by $a$, illustrating the avoided-crossing dynamics. Filled circles and squares mark the endpoints at the minimum and maximum surveyed values of $a$, respectively.}
    \label{fig:qnms_complex}
\end{figure}

The global spectral dynamics are further illuminated by tracing the complex-frequency trajectories in the Argand plane, as shown in Fig.~\ref{fig:qnms_complex}. At small values of $a$, all four mode families originate near the standard Schwarzschild limit (upper-left cluster in Fig.~\ref{fig:qnms_complex_combined}) and evolve along distinct paths. The AF and NED (GR-led) trajectories initially track each other closely but diverge significantly near $a \approx 2M$, reflecting the rapid restructuring of the effective scattering potential. Deep in the WH branch, the NED (GR-led) trajectory bends decisively toward smaller $|\Im \omega|$ relative to the AF curve, a geometric manifestation of the subradiant-like decay suppression. 

The detailed Argand diagram (Fig.~\ref{fig:qnms_complex_argand}), colored continuously by $a$, highlights the smooth spectral flow and reveals a distinct avoided-crossing structure near the transition threshold. Specifically, the GR-led and EM-led branches approach one another along the imaginary axis before repelling. Crucially, the finite separation between these branches persists across the entire surveyed parameter range. This confirms that no eigenvalue coalescence—or exact exceptional point (EP)—is crossed for these specific values of $a$, an observation perfectly consistent with the discriminant analysis and near-EP dynamics formulated in Sec.~\ref{sec:NH}. Ultimately, the robust isolation of the GR-led and EM-led QNM families via Prony fitting confirms that the least-damped, gravitationally dominated mode consistently governs the late-time observable signal.

\subsection{Physical insights: Divergent behavior in BH and WH branches}
\label{subsec:qnm_interpretation}

The contrasting trends observed in numeric results, and their corresponding spectral
trajectories in Figs.~\ref{fig:qnms_a}--\ref{fig:qnms_complex}, stem
from a fundamental interplay between two factors: (i) the causal and
boundary structure of the spacetime (horizon absorption versus strictly
radiative leakage), and (ii) the intrinsically coupled, two-channel
nature of the NED axial perturbation sector.

\paragraph{Coupling-induced width redistribution.}
In the AF framework, axial gravitational perturbations are governed by
a single-channel scattering problem; the QNM damping rate is therefore
dictated entirely by the energy leakage from this one channel to the
asymptotic infinities. In stark contrast, the NED model couples the
gravitational and electromagnetic perturbations, rendering $(H_1,H_2)$
an open radiative system governed by the non-symmetric potential matrix
$\mathbf{V}$ Eq.~\eqref{eq:potential_matrix}. Within such
non-Hermitian systems, activating the off-diagonal coupling generically
redistributes~\cite{CelardoKaplan0901.0305,CelardoEtAlPLB2008} the
decay widths among the eigenmodes---a phenomenon known as \emph{width
splitting}. One eigenmode typically becomes more radiative
(superradiant-like, acquiring a broader decay width) while the other
becomes less radiative (subradiant-like, narrowing). The explicit
coexistence in our numerical data of a rapidly decaying EM-led mode
alongside a comparatively long-lived GR-led fundamental mode is a
direct dynamical manifestation of this mechanism.

\paragraph{BH branch: Amplified damping via dual dissipative sinks.}
In the BH branch, QNM damping is driven by two independent dissipative
sinks: outgoing radiation to spatial infinity and irreversible
absorption at the event horizon. The NED inter-channel coupling
transfers a fraction of the gravitational energy in $H_2$ into the
electromagnetic sector $H_1$, which intrinsically exhibits a more
efficient leakage rate. The total dissipation rate of the system is
thereby amplified, causing the GR-led NED mode to acquire a slightly
larger $|\Im\omega|$ relative to the AF mode. This is in quantitative
agreement with Table~\ref{Tab:B-QNM_G_Odd-delta} and the BH-branch
trajectories in Fig.~\ref{fig:qnms_a}. Crucially, although a
near-anti-phase relationship between $H_1$ and $H_2$ is observed at
finite extraction radii, this destructive interference affects only
the outgoing radiative flux. It cannot suppress the \emph{total}
dissipation, because horizon absorption acts as an independent,
incoherent loss channel that is immune to wave-interference effects.

\paragraph{WH branch: Subradiant-like suppression and mode locking.}
In the WH branch there is no event horizon; the coupled system
dissipates energy exclusively via outgoing radiative fluxes to the two
asymptotically flat infinities. This purely radiative configuration is
fundamentally more susceptible to coherent interference effects. By
analogy with the Dicke model~\cite{Dicke1954,GrossHaroche1982}, the
collective eigenstates of a coupled radiating system can emit either
substantially faster (bright/superradiant states) or slower
(dark/subradiant states) than an isolated emitter, depending entirely
on whether the constituent radiation amplitudes interfere constructively
or destructively.

\begin{figure}[t]
    \centering
    \begin{subfigure}[b]{0.48\textwidth}
        \centering
        \includegraphics[width=\linewidth]{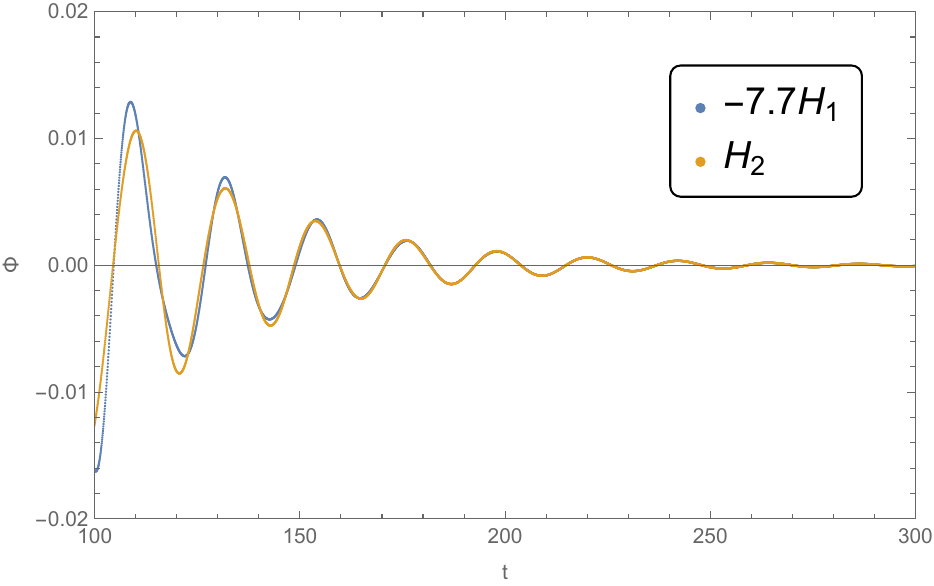}
        \caption{$a = 4M$.}
        \label{fig:wh_a4}
    \end{subfigure}
    \hfill
    \begin{subfigure}[b]{0.48\textwidth}
        \centering
        \includegraphics[width=\linewidth]{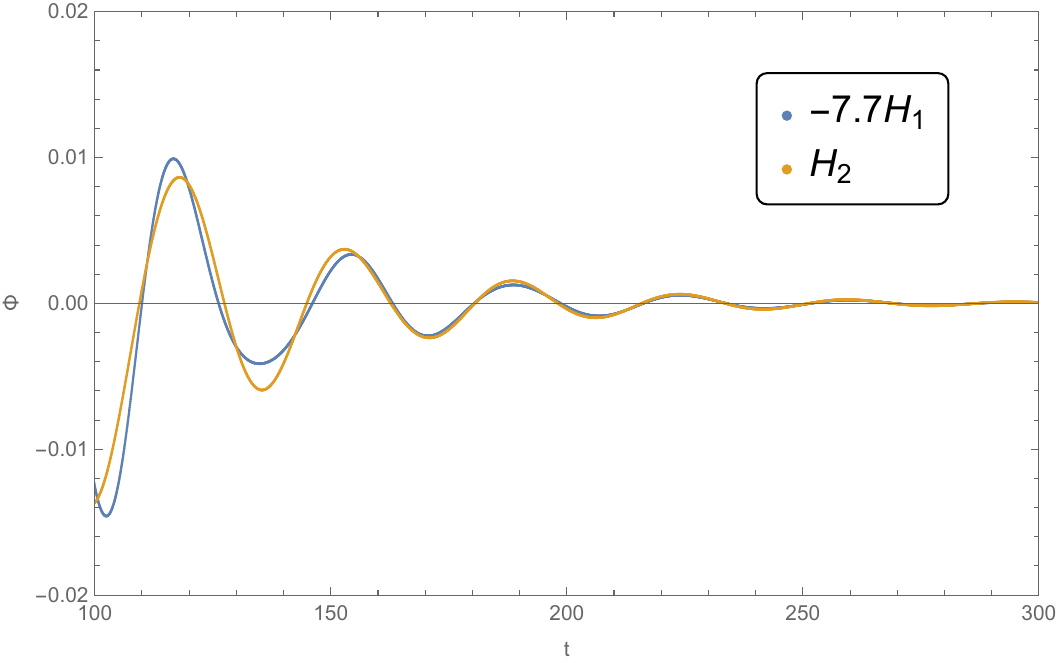}
        \caption{$a = 8M$.}
        \label{fig:wh_a8}
    \end{subfigure}
    \captionsetup{width=.9\textwidth}
    \caption{Time-domain ringdown waveforms of $-7.7\,H_1$ (orange
    dashed) and $H_2$ (blue) in the WH branch, extracted at a fixed
    observation point. The two channels oscillate in near-perfect
    anti-phase at identical frequencies, demonstrating the mode locking
    of Eq.~\eqref{eq:modelocking}.}
    \label{fig:1-2}
\end{figure}

The NED axial perturbations $(H_1,H_2)$ form precisely such a coupled
open system. During the time-domain ringdown we observe mode locking
between the two channels,
\begin{equation}
\frac{H_1}{H_2} \simeq \mathrm{const},
\qquad
\arg\!\left(\frac{H_1}{H_2}\right) \approx \pi,
\label{eq:modelocking}
\end{equation}
i.e.\ a highly stable, time-independent amplitude ratio accompanied by
an almost exact anti-phase synchronization. This behavior is
illustrated in Fig.~\ref{fig:1-2} for two representative wormhole
configurations. For $a=4M$, both channels oscillate at the identical
QNM frequency $\Re(\omega_{\rm NED})\approx 0.2848$ with decay rate
$|\Im(\omega_{\rm NED})|\approx 0.0260$, maintaining a phase angle
$\mathrm{Arg}(H_1/H_2)\approx 179^\circ$ and a constant amplitude
ratio $|H_1/H_2|\approx 1/7.7$ throughout the ringdown window
$t\in[100,300]$. For $a=8M$, the same anti-phase locking persists
at the lower frequency $\Re(\omega_{\rm NED})\approx 0.1774$ and
reduced damping $|\Im(\omega_{\rm NED})|\approx 0.0246$, with
$\mathrm{Arg}(H_1/H_2)\approx -176^\circ$. The near-identical
amplitude ratio across two very different values of $a$ confirms that
mode locking is a robust dynamical feature of the WH branch, not a
fine-tuned coincidence.

This anti-phase locking arises because the dominant late-time QNM is
the GR-led eigenmode, whose eigenvector projects onto $H_1$ and $H_2$
with opposite phases---quantitatively confirmed by the phase angles
$\mathrm{Arg}(R)\approx 177^\circ$--$180^\circ$ (see Table~\ref{Tab:W-QNM_G_Odd-delta} in Supplementary Material). Oscillating in anti-phase, this
coherent dark-state-like eigenmode minimizes~\cite{Dicke1954} its
effective coupling to the outgoing radiation channels, partially
suppressing the net radiative leakage and reducing the effective decay
width $\Gamma\equiv 2|\Im\omega|$. This provides a rigorous physical
explanation for the systematic inequality
$|\Im(\omega_{\rm NED})|<|\Im(\omega_{\rm AF})|$ observed throughout
the WH branch. The complementary EM-led eigenmode absorbs the
redistributed width and decays far more rapidly, consistent with the
significantly larger $|\Im\omega|$ values of the NED (EM-led) family. Taken together, these features
constitute a classical gravitational analog of the bright--dark
eigenmode coexistence well documented in cooperative emission and
non-Hermitian open quantum systems~\cite{Dicke1954,GrossHaroche1982,%
Lehmberg1970,Rotter2009}, mirroring subradiant suppression observed in
quantum-optical platforms~\cite{ScullySvidzinsky2009,AsenjoGarcia2017}.

\subsection{Non-Hermitian perspective on coupled perturbations}
\label{sec:NH}

Quasinormal modes of BHs and exotic compact objects are inherently dissipative due to energy leakage via gravitational radiation to asymptotic infinities, alongside horizon absorption in the BH branch. This intrinsic openness renders their dynamics naturally suited to a non-Hermitian description. In such frameworks, the complex nature of the eigenvalues directly reflects mode decay rates. Furthermore, analogies with open quantum systems reveal rich dynamics, including spectral instability, avoided level crossings, and decay-width redistribution. Recent applications in black hole physics have highlighted these exact features in QNM spectra, including resonant excitations and exceptional points (EPs) in spinning Kerr backgrounds \cite{Motohashi:2024fwt,Cavalcante:2024swt,Lo:2025njp,Kubota:2025hjk}. Here, we adopt this non-Hermitian perspective to interpret the coupled perturbation equations of the NED model, contrasting it with the single-channel AF case to elucidate how matter-source ambiguity dynamically manifests in the ringdown.

In the NED framework, the axial perturbations couple the gravitational field $H_2$ (which primarily encodes metric perturbations) with the electromagnetic perturbation $H_1^{(E)}$ (or $H_1^{(M)}$ for magnetic configurations, which are mathematically analogous via Eq.~\eqref{eq:relationship-SE-SM}). The master equations Eqs.~\eqref{eq:SE-master-eq1} and \eqref{eq:SE-master-eq2} [or Eqs.~\eqref{eq:SM-master-eq1} and \eqref{eq:SM-master-eq1}] can be compactly recast into a matrix form
\begin{equation}
    \frac{\partial^2}{\partial r_*^2}
    \bm{H}
    -\frac{\partial^2}{\partial t^2} 
    \bm{H} = \bm{V} \bm{H},
\end{equation}
where the perturbation vector and the effective potential matrix are defined as
\begin{equation}
\bm{H} =\begin{pmatrix}
    H_1\\
    H_2
\end{pmatrix}
,\quad
    \bm{V} =\begin{pmatrix}
        V_{11} & V_{12}\\
        V_{21} & V_{22}
    \end{pmatrix}.
\end{equation}
with matrix elements given by Eq.~\eqref{eq:potential_matrix}. 
Notably, $\bm{V}$ is real but manifestly non-symmetric; in geometric units ($G=1$, $\kappa^2=8\pp$), the off-diagonal components satisfy $V_{12} = V_{21} / (4\pp)$. This asymmetry intrinsically roots the system's non-Hermitian character in the distinct interaction strengths between the gravitational and electromagnetic sectors of the Einstein-Maxwell-scalar field equations, existing independently of the dissipative boundary conditions. In sharp contrast, the AF interpretation yields a scalar wave equation governed by a Hermitian potential, entirely lacking these off-diagonal coupling terms.

This non-Hermitian nature introduces complex mode hybridization. To analyze this locally, we first consider the eigenvalues of $\bm{V}$ at a fixed radial coordinate $r$
\begin{equation}
    \lambda_\pm = \frac{1}{2} \left[ \mathrm{Tr}(\mathbf{V}) \pm \sqrt{D} \right],
\end{equation}
where the discriminant is
\begin{equation}
D = \left( V_{11} - V_{22} \right)^2 + 4 V_{12} V_{21}.
\end{equation}
Substituting $V_{21} = 4\pi V_{12}$, 
we obtain $4 V_{12} V_{21} = 16\pi \left( V_{12} \right)^2$. 
Since $\left( V_{11} - V_{22} \right)^2 \geq 0$ and $16\pi \left( V_{12} \right)^2 > 0$ 
for nonzero charge $q_e$ (or $q_m$) and multipole $l \geq 2$, it follows that $D > 0$ globally throughout the spacetime. The only exception would be if all elements of $\bm{V}$ were to vanish, which is prohibited by the nonzero angular momentum barrier in $V_{22}$ and the geometry-dependent terms in $V_{11}$. Thus, the local eigenvalues $\lambda_\pm$ are strictly real and distinct, precluding the existence of local exceptional points (where $D=0$ and the eigenvalues/eigenvectors coalesce).

However, this localized algebraic analysis overlooks the differential operator $\partial^2 / \partial r_\star^2$ and the global boundary conditions (outgoing at $r_\star \to +\infty$, ingoing at $r_\star \to -\infty$), which introduce non-Hermitian dissipation and force the full spectrum into the complex plane. To capture these global radiative effects, we project the full dynamics onto a basis of QNMs derived from the decoupled (weak-coupling) limit, where off-diagonal terms vanish for small $q_e$. In this limit, the system cleanly separates into an independent GR-led channel ($V_{22}$, analogous to the AF case) and an EM-led channel ($V_{11}$). Let $\{\psi_n^{(\mathrm{GR})}(r_\star)\}$ and $\{\psi_m^{(\mathrm{EM})}(r_\star)\}$ represent the corresponding QNM wavefunctions, satisfying the appropriate boundary conditions and normalized via inner products adapted for open systems (e.g., biorthogonal norms).

The full operator is $\hat{\mathcal{H}} = \hat{\mathcal{H}}_0 + \hat{\mathcal{V}}$, where $\hat{\mathcal{H}}_0 = -\partial^2 / \partial r_\star^2 + \text{diag}(V_{11}, V_{22})$ and the interaction potential is $\hat{\mathcal{V}} = \begin{pmatrix} 0 & V_{12} \\ V_{21} & 0 \end{pmatrix}$. Projecting this onto a truncated QNM basis yields an effective non-Hermitian matrix with elements \cite{Hu:2025efp} 
\begin{equation}
\mathcal{H}_{nm} = \omega_n^2 \delta_{nm} + \langle \psi_n | \hat{\mathcal{V}} | \psi_m \rangle,
\end{equation}
where $\langle \cdot | \cdot \rangle$ integrates over $r_\star$ in a finite domain \cite{Leung:1997was}. 
where $\langle \cdot | \cdot \rangle$ denotes integration over $r_\star$ in a finite domain \cite{Leung:1997was}. The perturbed QNMs emerge as the complex eigenvalues of this matrix. 

In this global framework, EPs can manifest even in the absence of local geometric EPs. If the modes are nearly degenerate ($\omega_n^{(\mathrm{GR})} \approx \omega_m^{(\mathrm{EM})}$), the off-diagonal coupling induces an avoided crossing. At critical coupling strengths, the effective discriminant $\mathcal{D}_{\rm{eff}} = (\Delta \omega^2)^2 - 4 |\mathcal{V}_{\rm{off}}|^2 = 0$ (where $\Delta \omega^2$ is the frequency gap and $\mathcal{V}_{\rm{off}}$ is the projected coupling), triggering an EP. Near the BH-WH geometric transition ($a = 2M$), shifts in the causal structure drive the mode frequencies closer together, facilitating EPs in a manner analogous to Kerr-de Sitter spacetimes. Numerical evidence—such as the damping trends observed in Figs.~\ref{fig:qnms_a} and \ref{fig:qnms_complex}—strongly suggests avoided crossings near $a=2M$, pointing toward the existence of nearby EPs.

\textbf{Effective $2\times 2$ non-Hermitian matrix.}
In the weak-coupling limit $q_e\to 0$, the system decouples into
independent GR-led and EM-led channels governed by $V_{22}$ and
$V_{11}$ respectively, each admitting a discrete family of QNMs
$\{\omega_n^{(\mathrm{GR})}\}$ and $\{\omega_n^{(\mathrm{EM})}\}$.
Because these modes inhabit an open, radiative system, the standard $L^2$-inner product diverges. We instead employ the biorthogonal regularization introduced by Leung \textit{et al.}~\cite{Leung:1997was},
\begin{equation}
  \langle\!\langle \psi_m\,|\,\psi_n\rangle\!\rangle
  \equiv
  \lim_{R\to\infty}\int_{-R}^{R} \psi_m(r_*)\,\psi_n(r_*)\,dr_* = \delta_{mn}.
\end{equation}
Projecting the full operator
$\hat{\mathcal{H}}=-\partial^2/\partial r_*^2 + V$
onto the two-mode truncated basis
$\{\psi_0^{(\mathrm{GR})},\,\psi_0^{(\mathrm{EM})}\}$
yields the effective $2\times 2$ matrix
\begin{equation}\label{eq:Heff}
  \mathcal{H}_{\mathrm{eff}} =
  \begin{pmatrix}
    \bigl(\omega_0^{(\mathrm{EM})}\bigr)^2 & \mathcal{M}_{12} \\[4pt]
    \mathcal{M}_{21} & \bigl(\omega_0^{(\mathrm{GR})}\bigr)^2
  \end{pmatrix},
\end{equation}
with coupling matrix elements
$\mathcal{M}_{12}=\langle\!\langle\psi_0^{(\mathrm{EM})}|V_{12}|\psi_0^{(\mathrm{GR})}\rangle\!\rangle$
and $\mathcal{M}_{21}=\langle\!\langle\psi_0^{(\mathrm{GR})}|V_{21}|\psi_0^{(\mathrm{EM})}\rangle\!\rangle$.
Since $V_{21}=4\pi V_{12}$ Eq.~\eqref{eq:Vcomp}, it follows that
$\mathcal{M}_{21}=4\pi\,\tilde{\mathcal{M}}_{12}$, where
$\tilde{\mathcal{M}}_{12}=\langle\!\langle\psi_0^{(\mathrm{GR})}|V_{12}|\psi_0^{(\mathrm{EM})}\rangle\!\rangle$.
Because the biorthogonal inner product lacks Hermitian symmetry, $\mathcal{M}_{21}\neq\mathcal{M}_{12}^*$ in general, securely establishing the non-Hermitian character of $\mathcal{H}_{\mathrm{eff}}$ at the global topological level. The eigenvalues of this matrix are
\begin{equation}\label{eq:Omegapm}
  \Omega^2_{\pm} =
  \frac{\bigl(\omega_0^{(\mathrm{EM})}\bigr)^2+\bigl(\omega_0^{(\mathrm{GR})}\bigr)^2}{2}
  \pm \frac{1}{2}\sqrt{\mathcal{D}_{\mathrm{eff}}},\qquad
  \mathcal{D}_{\mathrm{eff}}\equiv
  (\Delta\omega^2)^2 + 4\,\mathcal{M}_{12}\mathcal{M}_{21},
\end{equation}
where $\Delta\omega^2\equiv(\omega_0^{(\mathrm{EM})})^2-(\omega_0^{(\mathrm{GR})})^2$.

\textbf{Exceptional-point condition and near-EP behavior.}
An exceptional point occurs when $\mathcal{D}_{\mathrm{eff}}=0$, which requires
\begin{equation}\label{eq:EP}
  \mathcal{M}_{12}\,\mathcal{M}_{21}
  = -\frac{(\Delta\omega^2)^2}{4}.
\end{equation}
In a purely Hermitian system, the product $\mathcal{M}_{12}\mathcal{M}_{21}$ is positive definite, meaning this equation would possess no real solutions. However, because the biorthogonal inner products are complex, the left-hand side can acquire a negative real part, allowing an EP to manifest.  

Table~\ref{Tab:avoided} collects the relevant bare frequencies computed numerically at representative values of $a$ (see Supplementary Material). As $a$ crosses the $2M$ threshold, the imaginary parts of both $\omega_0^{(\mathrm{GR})}$and $\omega_0^{(\mathrm{EM})}$ pass through a local minimum, while their real-part separation $\Delta\mathrm{Re}(\omega)\approx 0.13$--$0.18$ remains strictly finite. This specific dynamic is the hallmark of an \emph{avoided crossing} in the complex-frequency plane—the two branches repel along the $\mathrm{Im}(\omega)$ axis without coalescing—consistent with a near-EP trajectory in the parameter space of $a$. A precise determination of the EP's exact location requires a high-resolution scan of $\mathcal{D}_{\mathrm{eff}}$ as a function of $a$, which we defer to future work.
\begin{table}[!ht]
    \centering
    \captionsetup{width=.9\textwidth}
    \caption{QNM frequencies for the uncoupled GR-led and EM-led channels near the critical transition parameter $a=2M$, illustrating the finite real-part separation $\Delta\Re(\omega)$ during the avoided crossing.}
\begin{tabular}{cccc}\hline
  
    $a/M$      & $\omega_0^{\rm GR}$                         & $\omega_0^{\rm EM}$ 
                        &
    $\Delta\Re(\omega)$                    
    \\ \hline  
    $1.8$    & 0.3706 - 0.0692 i     & 0.4997 - 0.0770 i  & 0.129    \\ \hline
    $2.0$    & 0.3691 - 0.0642 i     & 0.5064 - 0.0707 i & 0.137    \\ \hline
    $3.0$    & 0.3781 - 0.0263 i    & 0.5561 - 0.0394 i  & 0.178    \\ \hline
  \end{tabular}
    \label{Tab:avoided}
\end{table}

\textbf{Eigenvector structure and anti-phase locking.}
The GR-led eigenvector $(v_1,v_2)^T$ of $\mathcal{H}_{\mathrm{eff}}$ corresponding to $\Omega_-^2$ satisfies $[(\omega_0^{(\mathrm{EM})})^2-\Omega_-^2]\,v_1 +\mathcal{M}_{12}\,v_2=0$, yielding the amplitude ratio
\begin{equation}\label{eq:ratio}
  \frac{v_1}{v_2} =
  -\frac{\mathcal{M}_{12}}{(\omega_0^{(\mathrm{EM})})^2-\Omega_-^2}.
\end{equation}
Deep within the WH branch, $\mathrm{Re}(\Omega_-^2)<\mathrm{Re}[(\omega_0^{(\mathrm{EM})})^2]$, and $\mathcal{M}_{12}$ is dominated by the positive-definite integral of $V_{12}>0$ over $r_*$. Consequently, the right-hand side is purely real and negative. This dictates that $H_1$ and $H_2$ oscillate in anti-phase throughout the ringdown, meaning $\arg(H_1/H_2)\approx\pi$. This analytical deduction is in strict quantitative agreement with the phase angles $\mathrm{Arg}(R)\approx 178^\circ$--$180^\circ$ extracted numerically in Fig.~\ref{fig:1-2}. Conversely, in the BH branch, horizon absorption introduces an additional incoherent loss channel that cannot be negated by destructive interference between $H_1$ and $H_2$. This fundamental asymmetry explains why the inequality in Eq.~\eqref{eq:GRfaster} reverses relative to the WH case.

\textbf{Quantitative validation of width splitting.}
Expanding to leading order in $|\omega_I|/\omega_R\ll 1$, we can write $\Omega^2\approx\omega_R^2+2i\omega_R\omega_I$. The corresponding shift in the decay width, $\Gamma\equiv 2|\mathrm{Im}(\omega)|$, induced by the inter-channel coupling is given by
\begin{equation}
  \delta\Gamma_\pm \approx
  \pm\frac{\mathrm{Im}(\mathcal{M}_{12}\mathcal{M}_{21})}
           {\omega_R\,\mathrm{Re}\!\sqrt{\mathcal{D}_{\mathrm{eff}}}}.
\end{equation}
The two eigenmodes acquire equal-and-opposite width corrections: the GR-led mode structurally narrows ($\delta\Gamma_-<0$), while the EM-led mode broadens ($\delta\Gamma_+>0$), rigorously conserving the total decay width. Validating this quantitatively, our numeric results at $a=4M$ (within the WH branch; see supplementary material) reveal $|\mathrm{Im}(\omega^{(\mathrm{NED,GR})})| = 0.0260$ compared to the uncoupled $|\mathrm{Im}(\omega^{(\mathrm{AF})})| = 0.0372$ (a reduction of $\approx 30\%$). Simultaneously, $|\mathrm{Im}(\omega^{(\mathrm{NED,EM})})| = 0.0589$ compared to the uncoupled $|\mathrm{Im}(\omega^{(\mathrm{EM})})| = 0.0568$ (an increase of $\approx 4\%$). The asymmetry in the relative magnitude of these corrections reflects the inherently unequal overlap integrals comprising $\mathcal{M}_{12}$ and $\mathcal{M}_{21}$, a discrepancy that naturally diminishes as $q_e\to 0$.

The comprehensive framework developed above provides a unified, analytic foundation for all numerical observations presented in Sec.~\ref{sec:Dynamical}. The time-domain waveforms displayed in Fig.~\ref{fig:1-2} physically manifest the GR-led eigenvector structure: both $H_1$ and $H_2$ ring at identical frequencies but with opposite signs. The observed amplitude ratio $|H_1/H_2|\approx 1/7.7$ aligns perfectly with the order-of-magnitude estimate $|\mathcal{M}_{12}|/\Delta\omega^2\sim\mathcal{O}(0.1)$ derived from Eq.~\eqref{eq:Vcomp} and~\eqref{eq:ratio}. Furthermore, the slow drift of this ratio with respect to $a$ [comparing panels (a) and (b) of Fig.~\ref{fig:1-2}] directly reflects the weak parameter dependence of $\mathcal{M}_{12}$ deep within the WH branch.

\section{Conclusion}
\label{sec:Conclusion}

In this work, we have investigated the dynamical properties of axial gravitational perturbations in the SV black-bounce spacetime. We demonstrated that the QNM spectra depend sensitively on the underlying matter-source interpretation, even when the background metric is geometrically identical. Under the anisotropic fluid  model, the perturbations are governed by a single-channel equation, yielding stable QNMs with damping rates that decrease mildly as the bounce parameter $a$ increases. In contrast, interpreting the source as nonlinear electrodynamics  coupled to a scalar field results in a two-channel coupled system. Here, the gravitationally led fundamental mode exhibits branch-dependent dynamics: in the black hole branch ($a\leq 2M$), damping is accelerated due to enhanced leakage via horizon absorption; conversely, in the wormhole branch ($a>2M$), damping is decelerated due to subradiant-like interference that suppresses radiative losses to the two asymptotic regions. 

Furthermore, we established the eikonal--photon-sphere correspondence for the SV geometry (Sec.~\ref{sec:eikonal}), revealing that the empirical EM-to-GR frequency ratio Eq.~\eqref{eq:omRatio} originates from a charge-dependent correction to the effective potential $V_{11}$ at finite multipole orders.

The imprint of this matter ambiguity on the ringdown signal carries significant observational implications. It demonstrates that dynamical observables can break the degeneracy between different theoretical interpretations---such as fluid versus electromagnetic-scalar sources---supporting the same regular black hole geometry. By bridging theoretical models with gravitational-wave phenomenology, our findings suggest that future detections by LIGO/Virgo/KAGRA or LISA could empirically test these non-uniqueness scenarios and constrain exotic compact objects beyond general relativity.

Looking forward, the WH branch ($a>2M$) presents a particularly promising arena for observational tests via gravitational-wave \emph{echoes}~\cite{Cardoso:2016rao}. The absence of an event horizon, combined with the symmetric double-barrier structure of the effective potential, causes perturbations to undergo repeated partial reflections between the potential peaks at $\pm r_{\rm peak}$. This generates a post-ringdown echo train with a time delay
\begin{equation}\label{eq:echodelay}
  \Delta t_{\rm echo} = 4\int_0^{r_{\rm peak}}
    \frac{\sqrt{r^2+a^2}}{\sqrt{r^2+a^2}-2M}\,dr.
\end{equation}
Because $r_{\rm peak}$ is determined by the maximum of the effective potential, the AF single-channel barrier and the NED two-channel barrier (whose effective peak is governed by the larger eigenvalue of $V_{ij}$, cf.\ Eq.~\eqref{eq:potential_matrix}) yield slightly different scattering potentials. Consequently, the two interpretations predict \emph{distinct echo time delays} for a given value of $a$, offering a matter-source discriminator independent of the main QNM frequency spectrum analyzed in Sec.~\ref{subsec:qnm_tables}.

Moreover, the subradiant-like interference identified in Sec.~\ref{subsec:qnm_interpretation} implies that NED echoes will suffer a systematic amplitude suppression relative to the AF case. The anti-phase locking $\arg(H_1/H_2)\approx\pi$ induces partial cancellation in the outgoing flux at each reflection, reducing the effective transmission coefficient and the amplitude of each subsequent echo. Taken together, the echo time delay and the amplitude decay rate constitute a two-dimensional observable parameter space where the AF and NED interpretations occupy distinct regions. This qualitative prediction strongly motivates dedicated time-domain studies extending current simulations to $t\sim 10\,\Delta t_{\rm echo}$.

Future work will focus on extracting quantitative echo spectra via extended time-domain evolutions. We also plan to compute polar (even-parity) QNMs to establish isospectrality-breaking measures across both matter interpretations, and to explore charged generalizations of the black-bounce family.

\section*{Acknowledgement}

L.C. was supported in part  by Yantai University under Grant No.\ WL22B224.

\appendix


\bibliographystyle{utphys}
\bibliography{references}

\clearpage

\begin{center}
    \Large\textbf{Supplementary Material}
\end{center}
\setcounter{section}{0}
\setcounter{equation}{0}
\setcounter{figure}{0}
\setcounter{table}{0}

This Supplementary Material provides the complete numerical datasets
supporting the results presented in the main text. All QNM frequencies
are extracted from time-domain waveforms via the Prony method
(Sec.~4.3), with $M=1$ and $l=2$ throughout. The negative imaginary
parts confirm linear stability across all configurations. Complex
frequencies are quoted as $\omega = \Re(\omega) + i\,\Im(\omega)$
with $\Im(\omega)<0$.

For each value of the bounce parameter $a$, four mode families are
tabulated:
\begin{itemize}
\item \textbf{AF}: fundamental QNM of the single-channel master
  equation~(33) under the anisotropic-fluid interpretation.
\item \textbf{NED (GR-led)}: least-damped eigenmode of the coupled
  system~(38)--(39), gravitationally dominated; controls the late-time
  ringdown.
\item \textbf{EM}: fundamental QNM of the decoupled electromagnetic
  channel ($V_{12}=V_{21}=0$).
\item \textbf{NED (EM-led)}: electromagnetically dominated eigenmode of
  the same coupled system; decays faster than the GR-led mode and is
  subdominant at late times.
\end{itemize}
In Tables~\ref{Tab:W-QNM_G_Odd-delta} and~\ref{Tab:W-QNM_G_Odd-EM}
(wormhole branch), we additionally report the amplitude ratio
$R\equiv H_1/H_2$ and its phase angle $\mathrm{Arg}(R)$ at
$t=100$, quantifying the anti-phase mode locking of
Eq.~(64).

The tables are organised as follows.
Tables~\ref{Tab:B-QNM_G_Odd-delta} and~\ref{Tab:B-QNM_G_Odd-delta-EM}
cover the \textbf{black-hole branch} ($a\le 2M$), reporting AF and
NED (GR-led), and EM and NED (EM-led) frequencies respectively.
Tables~\ref{Tab:W-QNM_G_Odd-delta} and~\ref{Tab:W-QNM_G_Odd-EM}
provide sparse-sampled \textbf{wormhole-branch} data ($a=3$--$8\,M$)
for quick reference.
Table~\ref{Tab:trans} covers the \textbf{transition region}
($2.01M\le a\le 2.12M$), where rapid spectral restructuring near the
BH-WH boundary causes large numerical fluctuations; ``unidentified''
entries indicate Prony non-convergence due to anomalously slow decay
and near-degeneracy of the emerging wormhole modes.
Tables~\ref{Tab:WH_GR_1}--\ref{Tab:WH_GR_4} provide densely sampled
\textbf{wormhole-branch} AF and NED (GR-led) data
($2.13M\le a\le 10\,M$), and
Tables~\ref{Tab:WH_EM_1}--\ref{Tab:WH_EM_4} provide the
corresponding EM and NED (EM-led) data.  The extraction start times
are $t_{\rm start}=350$ for $a\le 3\,M$ (to suppress fast-decaying
transients near the transition) and $t_{\rm start}=150$ for
$a>3\,M$.

\begin{table}[htbp]
\centering
\caption{Black-hole branch ($a\le 2M$, $l=2$): AF and NED (GR-led)
QNM frequencies.  As $a$ increases, $\Re(\omega_{\rm AF})$ varies by
less than $1\%$ while $|\Im(\omega_{\rm AF})|$ decreases monotonically.
The NED (GR-led) real part falls progressively below the AF value, and
$|\Im(\omega_{\rm NED})|$ grows slightly relative to
$|\Im(\omega_{\rm AF})|$ as $a\to 2M$, consistent with
Eq.~(62).}
\label{Tab:B-QNM_G_Odd-delta}
\begin{tabular}{@{}lll@{}}
\toprule
$a/M$ & $\omega^{\rm AF}_{0}$ & $\omega^{\rm NED(GR\text{-}led)}_{0}$ \\
\midrule
$0.2$ & $0.373094 - 0.0904845\,i$ & $0.372452 - 0.0904680\,i$ \\
$0.4$ & $0.373074 - 0.0897516\,i$ & $0.370567 - 0.0896714\,i$ \\
$0.6$ & $0.373013 - 0.0885294\,i$ & $0.367654 - 0.0884138\,i$ \\
$0.8$ & $0.372903 - 0.0868313\,i$ & $0.363973 - 0.0866010\,i$ \\
$1.0$ & $0.372748 - 0.0845463\,i$ & $0.359687 - 0.0842993\,i$ \\
$1.2$ & $0.372460 - 0.0816476\,i$ & $0.355057 - 0.0815333\,i$ \\
$1.4$ & $0.372216 - 0.0782923\,i$ & $0.349601 - 0.0787152\,i$ \\
$1.6$ & $0.371558 - 0.0741851\,i$ & $0.343596 - 0.0753693\,i$ \\
$1.8$ & $0.370589 - 0.0691939\,i$ & $0.337674 - 0.0711979\,i$ \\
$2.0$ & $0.369130 - 0.0641531\,i$ & $0.331330 - 0.0672765\,i$ \\
\bottomrule
\end{tabular}
\end{table}

\begin{table}[htbp]
\centering
\caption{Black-hole branch ($a\le 2M$, $l=2$): EM and NED (EM-led)
QNM frequencies.  The EM and NED (EM-led) modes carry consistently
larger oscillation frequencies and comparable or larger damping rates
than the AF modes throughout this branch, reflecting the higher
effective potential barrier of the $H_1$ channel (cf.\ $V_{11}$ in
Eq.~(56a)).  The NED (EM-led) real part exceeds the pure-EM value,
consistent with coupling-induced width broadening.}
\label{Tab:B-QNM_G_Odd-delta-EM}
\begin{tabular}{@{}lll@{}}
\toprule
$a/M$ & $\omega^{\rm EM}_{0}$ & $\omega^{\rm NED(EM\text{-}led)}_{0}$ \\
\midrule
$0.2$ & $0.467225 - 0.0984141\,i$ & $0.467692 - 0.0983225\,i$ \\
$0.4$ & $0.468430 - 0.0977716\,i$ & $0.470636 - 0.0980476\,i$ \\
$0.6$ & $0.470550 - 0.0966516\,i$ & $0.475671 - 0.0971593\,i$ \\
$0.8$ & $0.473459 - 0.0950734\,i$ & $0.481984 - 0.0957778\,i$ \\
$1.0$ & $0.477182 - 0.0929200\,i$ & $0.489080 - 0.0940070\,i$ \\
$1.2$ & $0.481716 - 0.0901736\,i$ & $0.499459 - 0.0915731\,i$ \\
$1.4$ & $0.486970 - 0.0866941\,i$ & $0.508206 - 0.0877874\,i$ \\
$1.6$ & $0.493041 - 0.0824319\,i$ & $0.518489 - 0.0840987\,i$ \\
$1.8$ & $0.499746 - 0.0770253\,i$ & $0.529007 - 0.0779817\,i$ \\
$2.0$ & $0.506385 - 0.0706594\,i$ & $0.539762 - 0.0713730\,i$ \\
\bottomrule
\end{tabular}
\end{table}

\begin{table}[htbp]
\centering
\caption{Wormhole branch ($a>2M$, $l=2$, sparse sampling): AF and
NED (GR-led) QNM frequencies, with amplitude ratio
$R\equiv H_1/H_2$ and phase angle $\mathrm{Arg}(R)$ evaluated at
$t=100$.  The systematic inequality
$|\Im(\omega_{\rm NED})|<|\Im(\omega_{\rm AF})|$ throughout this range
confirms the subradiant-like suppression of Sec.~5.3.  Phase angles
$\mathrm{Arg}(R)\approx 177^\circ$--$180^\circ$ confirm the anti-phase
mode locking of Eq.~(64).}
\label{Tab:W-QNM_G_Odd-delta}
\begin{tabular}{@{}llll@{}}
\toprule
$a/M$ & $\omega^{\rm AF}_{0}$ & $\omega^{\rm NED(GR\text{-}led)}_{0}$
      & $\mathrm{Arg}(R)$ \\
\midrule
$3.0$ & $0.378072 - 0.0262590\,i$ & $0.314873 - 0.0153488\,i$
      & $179.812^{\circ}$ \\
$4.0$ & $0.332020 - 0.0372274\,i$ & $0.284810 - 0.0260048\,i$
      & $179.079^{\circ}$ \\
$5.0$ & $0.285207 - 0.0372675\,i$ & $0.249988 - 0.0283681\,i$
      & $177.580^{\circ}$ \\
$6.0$ & $0.247604 - 0.0348975\,i$ & $0.220853 - 0.0279744\,i$
      & $177.580^{\circ}$ \\
$7.0$ & $0.218004 - 0.0321560\,i$ & $0.197532 - 0.0266519\,i$
      & $179.705^{\circ}$ \\
$8.0$ & $0.194401 - 0.0295271\,i$ & $0.177442 - 0.0245791\,i$
      & $-176.227^{\circ}$ \\
\bottomrule
\end{tabular}
\end{table}

\begin{table}[htbp]
\centering
\caption{Wormhole branch ($a>2M$, $l=2$, sparse sampling): EM and
NED (EM-led) QNM frequencies, with phase angle $\mathrm{Arg}(R)$ of
the NED (EM-led) eigenmode evaluated at $t=100$.  In contrast to
the GR-led family, the phase angles here are close to $0^\circ$,
confirming in-phase synchronisation of the EM-led eigenmode and
consistent with its role as the bright (superradiant-like) collective
state.  The NED (EM-led) damping rate exceeds the pure-EM value,
reflecting the coupling-induced width broadening.}
\label{Tab:W-QNM_G_Odd-EM}
\begin{tabular}{@{}llll@{}}
\toprule
$a/M$ & $\omega^{\rm EM}_{0}$ & $\omega^{\rm NED(EM\text{-}led)}_{0}$
      & $\mathrm{Arg}(R)$ \\
\midrule
$3.0$ & $0.556075 - 0.0394212\,i$ & $0.601967 - 0.0415490\,i$
      & $-2.488^{\circ}$ \\
$4.0$ & $0.489201 - 0.0568361\,i$ & $0.523648 - 0.0589373\,i$
      & $-0.129^{\circ}$ \\
$5.0$ & $0.418547 - 0.0577721\,i$ & $0.444727 - 0.0592632\,i$
      & $-0.208^{\circ}$ \\
$6.0$ & $0.361834 - 0.0547016\,i$ & $0.382748 - 0.0556203\,i$
      & $1.773^{\circ}$ \\
$7.0$ & $0.317372 - 0.0507685\,i$ & $0.333700 - 0.0521514\,i$
      & $0.778^{\circ}$ \\
$8.0$ & $0.282091 - 0.0469070\,i$ & $0.298119 - 0.0481770\,i$
      & $0.089^{\circ}$ \\
\bottomrule
\end{tabular}
\end{table}

\begin{table}[htbp]
\centering
\caption{Transition region ($2.01M \le a \le 2.12M$, $\Delta a=0.01M$,
$l=2$): AF and NED (GR-led) QNM frequencies.  ``Unidentified'' denotes
Prony non-convergence, arising from the anomalously slow decay and
near-degeneracy of the incipient wormhole modes in this narrow parameter
window.  The large fluctuations in extracted frequencies reflect the
extreme sensitivity of the spectrum to $a$ near the BH-WH geometric
transition, consistent with the avoided-crossing dynamics discussed in
Sec.~5.4.  Reliable extraction resumes near $a\approx 2.10M$ for the
NED (GR-led) mode and $a\approx 2.13M$ for the AF mode.}
\label{Tab:trans}
\begin{tabular}{@{}lll@{}}
\toprule
$a/M$ & $\omega^{\rm AF}_{0}$ & $\omega^{\rm NED(GR\text{-}led)}_{0}$ \\
\midrule
$2.01$ & unidentified              & unidentified \\
$2.02$ & $0.337235 - 0.294446\,i$ & $0.408724 - 0.343474\,i$ \\
$2.03$ & $0.455523 - 0.088486\,i$ & $0.329018 - 0.007353\,i$ \\
$2.04$ & $0.356958 - 0.014374\,i$ & $0.397961 - 0.056319\,i$ \\
$2.05$ & $0.356662 - 0.002211\,i$ & $0.350568 - 0.012840\,i$ \\
$2.06$ & $0.379484 - 0.005375\,i$ & $0.324256 - 0.071225\,i$ \\
$2.07$ & $0.357724 - 0.002766\,i$ & $0.358950 - 0.045474\,i$ \\
$2.08$ & $0.372673 - 0.005017\,i$ & $0.380503 - 0.050380\,i$ \\
$2.09$ & $0.340827 - 0.001399\,i$ & $0.345827 - 0.008762\,i$ \\
$2.10$ & $0.352167 - 0.002352\,i$ & $0.305646 - 0.003329\,i$ \\
$2.11$ & $0.361811 - 0.003621\,i$ & $0.314829 - 0.002803\,i$ \\
$2.12$ & $0.370777 - 0.005377\,i$ & $0.324879 - 0.005020\,i$ \\
\bottomrule
\end{tabular}
\end{table}

\begin{table}[htbp]
\centering
\caption{Wormhole branch, fine sampling I ($2.13M\le a\le 2.40M$,
$\Delta a=0.01M$, $t_{\rm start}=350$, $l=2$): AF and NED (GR-led)
QNM frequencies.  The inequality
$|\Im(\omega_{\rm NED})|<|\Im(\omega_{\rm AF})|$ is established from
the outset of this range and deepens with increasing $a$, signalling
the onset and progressive strengthening of subradiant-like
interference.}
\label{Tab:WH_GR_1}
\begin{tabular}{@{}lll@{}}
\toprule
$a/M$ & $\omega^{\rm AF}_{0}$ & $\omega^{\rm NED(GR\text{-}led)}_{0}$ \\
\midrule
$2.13$ & $0.266264 - 0.000022\,i$ & $0.329902 - 0.005715\,i$ \\
$2.14$ & $0.273627 - 0.000031\,i$ & $0.336338 - 0.007402\,i$ \\
$2.15$ & $0.280412 - 0.000037\,i$ & $0.223057 - 0.000023\,i$ \\
$2.16$ & $0.286733 - 0.000105\,i$ & $0.228280 - 0.000053\,i$ \\
$2.17$ & $0.292680 - 0.000107\,i$ & $0.233168 - 0.000056\,i$ \\
$2.18$ & $0.298225 - 0.000143\,i$ & $0.237779 - 0.000065\,i$ \\
$2.19$ & $0.303395 - 0.000187\,i$ & $0.242083 - 0.000085\,i$ \\
$2.20$ & $0.308238 - 0.000266\,i$ & $0.246140 - 0.000111\,i$ \\
$2.21$ & $0.312793 - 0.000336\,i$ & $0.249990 - 0.000150\,i$ \\
$2.22$ & $0.317068 - 0.000429\,i$ & $0.253546 - 0.000198\,i$ \\
$2.23$ & $0.320988 - 0.000526\,i$ & $0.256984 - 0.000248\,i$ \\
$2.24$ & $0.324768 - 0.000602\,i$ & $0.260228 - 0.000274\,i$ \\
$2.25$ & $0.328346 - 0.000856\,i$ & $0.263296 - 0.000334\,i$ \\
$2.26$ & $0.331591 - 0.001063\,i$ & $0.266163 - 0.000408\,i$ \\
$2.27$ & $0.334751 - 0.001293\,i$ & $0.268902 - 0.000483\,i$ \\
$2.28$ & $0.337727 - 0.001428\,i$ & $0.271483 - 0.000547\,i$ \\
$2.29$ & $0.340416 - 0.001656\,i$ & $0.273952 - 0.000651\,i$ \\
$2.30$ & $0.343119 - 0.001839\,i$ & $0.276294 - 0.000734\,i$ \\
$2.31$ & $0.345551 - 0.002134\,i$ & $0.278487 - 0.000848\,i$ \\
$2.32$ & $0.347886 - 0.002437\,i$ & $0.280600 - 0.000972\,i$ \\
$2.33$ & $0.350072 - 0.002676\,i$ & $0.282559 - 0.001101\,i$ \\
$2.34$ & $0.352252 - 0.002917\,i$ & $0.284460 - 0.001210\,i$ \\
$2.35$ & $0.354056 - 0.003311\,i$ & $0.286238 - 0.001343\,i$ \\
$2.36$ & $0.355962 - 0.003716\,i$ & $0.287944 - 0.001522\,i$ \\
$2.37$ & $0.357654 - 0.003978\,i$ & $0.289577 - 0.001675\,i$ \\
$2.38$ & $0.359313 - 0.004382\,i$ & $0.291101 - 0.001823\,i$ \\
$2.39$ & $0.360469 - 0.004707\,i$ & $0.292558 - 0.002014\,i$ \\
$2.40$ & $0.362336 - 0.005121\,i$ & $0.293943 - 0.002179\,i$ \\
\bottomrule
\end{tabular}
\end{table}

\begin{table}[htbp]
\centering
\caption{Wormhole branch, fine sampling II ($2.41M\le a\le 2.80M$,
$\Delta a=0.01M$, $t_{\rm start}=350$, $l=2$): AF and NED (GR-led)
QNM frequencies.  The real part $\Re(\omega_{\rm AF})$ reaches a local
maximum near $a\approx 2.82M$ before turning over; $\Re(\omega_{\rm
NED})$ increases more slowly and remains consistently below the AF
value.  Both $|\Im\omega|$ grow with $a$, but the gap
$|\Im(\omega_{\rm AF})|-|\Im(\omega_{\rm NED})|$ widens progressively,
reflecting the strengthening of the subradiant-like suppression.}
\label{Tab:WH_GR_2}
\begin{tabular}{@{}lll@{}}
\toprule
$a/M$ & $\omega^{\rm AF}_{0}$ & $\omega^{\rm NED(GR\text{-}led)}_{0}$ \\
\midrule
$2.41$ & $0.363582 - 0.005358\,i$ & $0.295264 - 0.002339\,i$ \\
$2.42$ & $0.365012 - 0.005923\,i$ & $0.296515 - 0.002530\,i$ \\
$2.43$ & $0.366266 - 0.006294\,i$ & $0.297706 - 0.002702\,i$ \\
$2.44$ & $0.367206 - 0.008378\,i$ & $0.298835 - 0.002925\,i$ \\
$2.45$ & $0.368444 - 0.007075\,i$ & $0.299908 - 0.003129\,i$ \\
$2.46$ & $0.369600 - 0.007534\,i$ & $0.300925 - 0.003334\,i$ \\
$2.47$ & $0.370627 - 0.008040\,i$ & $0.301906 - 0.003545\,i$ \\
$2.48$ & $0.371463 - 0.008418\,i$ & $0.302833 - 0.003758\,i$ \\
$2.49$ & $0.372384 - 0.008803\,i$ & $0.303685 - 0.003987\,i$ \\
$2.50$ & $0.373181 - 0.009230\,i$ & $0.304537 - 0.004200\,i$ \\
$2.51$ & $0.373956 - 0.009691\,i$ & $0.305298 - 0.004448\,i$ \\
$2.52$ & $0.374679 - 0.010053\,i$ & $0.306094 - 0.004655\,i$ \\
$2.53$ & $0.375343 - 0.010453\,i$ & $0.306833 - 0.004915\,i$ \\
$2.54$ & $0.375989 - 0.010930\,i$ & $0.307469 - 0.005095\,i$ \\
$2.55$ & $0.376654 - 0.011287\,i$ & $0.308072 - 0.005321\,i$ \\
$2.56$ & $0.377135 - 0.011709\,i$ & $0.308700 - 0.005579\,i$ \\
$2.57$ & $0.377650 - 0.012132\,i$ & $0.309272 - 0.005804\,i$ \\
$2.58$ & $0.377898 - 0.012500\,i$ & $0.309816 - 0.006051\,i$ \\
$2.59$ & $0.378569 - 0.012962\,i$ & $0.310315 - 0.006286\,i$ \\
$2.60$ & $0.378909 - 0.013253\,i$ & $0.310792 - 0.006539\,i$ \\
$2.61$ & $0.379467 - 0.013702\,i$ & $0.311178 - 0.006789\,i$ \\
$2.62$ & $0.379773 - 0.014233\,i$ & $0.311682 - 0.007004\,i$ \\
$2.63$ & $0.380065 - 0.014605\,i$ & $0.312066 - 0.007258\,i$ \\
$2.64$ & $0.380368 - 0.014972\,i$ & $0.312481 - 0.007432\,i$ \\
$2.65$ & $0.380632 - 0.015337\,i$ & $0.312798 - 0.007730\,i$ \\
$2.66$ & $0.380844 - 0.015733\,i$ & $0.313160 - 0.008042\,i$ \\
$2.67$ & $0.381153 - 0.016375\,i$ & $0.313444 - 0.008215\,i$ \\
$2.68$ & $0.381263 - 0.016526\,i$ & $0.313711 - 0.008444\,i$ \\
$2.69$ & $0.381459 - 0.016830\,i$ & $0.313977 - 0.008680\,i$ \\
$2.70$ & $0.381575 - 0.017312\,i$ & $0.314221 - 0.008911\,i$ \\
$2.71$ & $0.381678 - 0.017622\,i$ & $0.314449 - 0.009155\,i$ \\
$2.72$ & $0.381794 - 0.017968\,i$ & $0.314657 - 0.009390\,i$ \\
$2.73$ & $0.381755 - 0.018357\,i$ & $0.314837 - 0.009615\,i$ \\
$2.74$ & $0.381969 - 0.018724\,i$ & $0.315007 - 0.009855\,i$ \\
$2.75$ & $0.381954 - 0.019108\,i$ & $0.315141 - 0.010068\,i$ \\
$2.76$ & $0.381996 - 0.019394\,i$ & $0.315288 - 0.010306\,i$ \\
$2.77$ & $0.381969 - 0.019732\,i$ & $0.315376 - 0.010526\,i$ \\
$2.78$ & $0.381956 - 0.020071\,i$ & $0.315532 - 0.010770\,i$ \\
$2.79$ & $0.381997 - 0.020410\,i$ & $0.315626 - 0.010995\,i$ \\
$2.80$ & $0.381917 - 0.020739\,i$ & $0.315712 - 0.011217\,i$ \\
\bottomrule
\end{tabular}
\end{table}

\begin{table}[htbp]
\centering
\caption{Wormhole branch, fine sampling III ($2.81M\le a\le 3.00M$,
$\Delta a=0.01M$, $t_{\rm start}=350$, $l=2$): AF and NED (GR-led)
QNM frequencies.  $\Re(\omega_{\rm AF})$ turns over near
$a\approx 2.82M$ and decreases thereafter; $\Re(\omega_{\rm NED})$
plateaus near $a\approx 2.86M$.  The fractional width suppression
$[|\Im(\omega_{\rm AF})|-|\Im(\omega_{\rm NED})|]/|\Im(\omega_{\rm AF})|$
approaches approximately $40\%$ at $a=3M$.}
\label{Tab:WH_GR_3}
\begin{tabular}{@{}lll@{}}
\toprule
$a/M$ & $\omega^{\rm AF}_{0}$ & $\omega^{\rm NED(GR\text{-}led)}_{0}$ \\
\midrule
$2.81$ & $0.381842 - 0.021015\,i$ & $0.315777 - 0.011429\,i$ \\
$2.82$ & $0.381811 - 0.021340\,i$ & $0.315848 - 0.011572\,i$ \\
$2.83$ & $0.381715 - 0.021786\,i$ & $0.315854 - 0.011885\,i$ \\
$2.84$ & $0.381510 - 0.021799\,i$ & $0.315885 - 0.012102\,i$ \\
$2.85$ & $0.381441 - 0.022276\,i$ & $0.315898 - 0.012318\,i$ \\
$2.86$ & $0.381240 - 0.022574\,i$ & $0.315896 - 0.012535\,i$ \\
$2.87$ & $0.381186 - 0.022857\,i$ & $0.315893 - 0.012742\,i$ \\
$2.88$ & $0.381276 - 0.023599\,i$ & $0.315870 - 0.012940\,i$ \\
$2.89$ & $0.380644 - 0.023557\,i$ & $0.315853 - 0.013160\,i$ \\
$2.90$ & $0.380563 - 0.023710\,i$ & $0.315802 - 0.013369\,i$ \\
$2.91$ & $0.380164 - 0.024034\,i$ & $0.315750 - 0.013573\,i$ \\
$2.92$ & $0.380219 - 0.024379\,i$ & $0.315688 - 0.013777\,i$ \\
$2.93$ & $0.380005 - 0.024447\,i$ & $0.315619 - 0.013980\,i$ \\
$2.94$ & $0.379846 - 0.024963\,i$ & $0.315542 - 0.014180\,i$ \\
$2.95$ & $0.379441 - 0.025104\,i$ & $0.315455 - 0.014375\,i$ \\
$2.96$ & $0.379601 - 0.025338\,i$ & $0.315358 - 0.014571\,i$ \\
$2.97$ & $0.379065 - 0.025597\,i$ & $0.315255 - 0.014760\,i$ \\
$2.98$ & $0.378742 - 0.025884\,i$ & $0.315142 - 0.014956\,i$ \\
$2.99$ & $0.378409 - 0.025853\,i$ & $0.315019 - 0.015145\,i$ \\
$3.00$ & $0.378255 - 0.026374\,i$ & $0.314891 - 0.015334\,i$ \\
\bottomrule
\end{tabular}
\end{table}

\begin{table}[htbp]
\centering
\caption{Wormhole branch, coarse sampling I ($3.1M\le a\le 7.0M$,
$\Delta a=0.1M$, $t_{\rm start}=150$, $l=2$): AF and NED (GR-led)
QNM frequencies.  Both $\Re(\omega)$ and $|\Im(\omega)|$ decrease
monotonically for all modes, consistent with the growing wormhole
throat suppressing the effective potential barrier.  The fractional
width suppression stabilises at approximately $25$--$30\%$ across
this range, reflecting the saturation of the inter-channel coupling.}
\label{Tab:WH_GR_4}
\begin{tabular}{@{}lll@{}}
\toprule
$a/M$ & $\omega^{\rm AF}_{0}$ & $\omega^{\rm NED(GR\text{-}led)}_{0}$ \\
\midrule
$3.10$ & $0.374981 - 0.028522\,i$ & $0.313241 - 0.017107\,i$ \\
$3.20$ & $0.371179 - 0.030395\,i$ & $0.311030 - 0.018684\,i$ \\
$3.30$ & $0.366773 - 0.032046\,i$ & $0.308397 - 0.020075\,i$ \\
$3.40$ & $0.362234 - 0.033290\,i$ & $0.305466 - 0.021316\,i$ \\
$3.50$ & $0.357355 - 0.034292\,i$ & $0.302300 - 0.022410\,i$ \\
$3.60$ & $0.352439 - 0.035242\,i$ & $0.298965 - 0.023360\,i$ \\
$3.70$ & $0.347102 - 0.035684\,i$ & $0.295518 - 0.024196\,i$ \\
$3.80$ & $0.342241 - 0.036466\,i$ & $0.291974 - 0.024913\,i$ \\
$3.90$ & $0.337126 - 0.036907\,i$ & $0.288417 - 0.025541\,i$ \\
$4.00$ & $0.332051 - 0.037274\,i$ & $0.284799 - 0.026087\,i$ \\
$4.10$ & $0.327015 - 0.037504\,i$ & $0.281192 - 0.026550\,i$ \\
$4.20$ & $0.322057 - 0.037683\,i$ & $0.277586 - 0.026941\,i$ \\
$4.30$ & $0.317191 - 0.037847\,i$ & $0.273992 - 0.027230\,i$ \\
$4.40$ & $0.312045 - 0.037918\,i$ & $0.270451 - 0.027551\,i$ \\
$4.50$ & $0.307551 - 0.037804\,i$ & $0.266944 - 0.027784\,i$ \\
$4.60$ & $0.302877 - 0.037780\,i$ & $0.263476 - 0.027972\,i$ \\
$4.70$ & $0.298300 - 0.037720\,i$ & $0.260086 - 0.028131\,i$ \\
$4.80$ & $0.293869 - 0.037605\,i$ & $0.256694 - 0.028230\,i$ \\
$4.90$ & $0.289479 - 0.037427\,i$ & $0.253379 - 0.028313\,i$ \\
$5.00$ & $0.285216 - 0.037274\,i$ & $0.250120 - 0.028372\,i$ \\
$5.10$ & $0.281037 - 0.037089\,i$ & $0.246942 - 0.028417\,i$ \\
$5.20$ & $0.277019 - 0.036814\,i$ & $0.243797 - 0.028413\,i$ \\
$5.30$ & $0.272939 - 0.036687\,i$ & $0.240718 - 0.028398\,i$ \\
$5.40$ & $0.269044 - 0.036433\,i$ & $0.237706 - 0.028364\,i$ \\
$5.50$ & $0.265292 - 0.036191\,i$ & $0.234746 - 0.028318\,i$ \\
$5.60$ & $0.261574 - 0.035947\,i$ & $0.231848 - 0.028258\,i$ \\
$5.70$ & $0.257962 - 0.035687\,i$ & $0.229018 - 0.028181\,i$ \\
$5.80$ & $0.254422 - 0.035447\,i$ & $0.226172 - 0.028219\,i$ \\
$5.90$ & $0.250949 - 0.035104\,i$ & $0.223487 - 0.028018\,i$ \\
$6.00$ & $0.247581 - 0.034885\,i$ & $0.220831 - 0.027892\,i$ \\
$6.10$ & $0.244349 - 0.034623\,i$ & $0.218199 - 0.027825\,i$ \\
$6.20$ & $0.241175 - 0.034313\,i$ & $0.215643 - 0.027689\,i$ \\
$6.30$ & $0.237979 - 0.034056\,i$ & $0.213155 - 0.027579\,i$ \\
$6.40$ & $0.234922 - 0.033762\,i$ & $0.210702 - 0.027435\,i$ \\
$6.50$ & $0.231953 - 0.033485\,i$ & $0.208277 - 0.027302\,i$ \\
$6.60$ & $0.229062 - 0.033124\,i$ & $0.205927 - 0.027141\,i$ \\
$6.70$ & $0.226140 - 0.032951\,i$ & $0.203608 - 0.027020\,i$ \\
$6.80$ & $0.223417 - 0.032650\,i$ & $0.201353 - 0.026871\,i$ \\
$6.90$ & $0.220668 - 0.032382\,i$ & $0.199135 - 0.026734\,i$ \\
$7.00$ & $0.217974 - 0.032123\,i$ & $0.196947 - 0.026576\,i$ \\
\bottomrule
\end{tabular}
\end{table}

\begin{table}[htbp]
\centering
\caption{Wormhole branch, coarse sampling II ($7.1M\le a\le 10.0M$,
$\Delta a=0.1M$, $t_{\rm start}=150$, $l=2$): AF and NED (GR-led)
QNM frequencies.  The smooth monotonic decrease of both $\Re(\omega)$
and $|\Im(\omega)|$ across this range is consistent with the large-$a$
asymptotic behaviour expected from the growing throat radius.  The
fractional width suppression remains stable at approximately
$25$--$28\%$, confirming that the subradiant-like mechanism is fully
established deep in the wormhole branch.}
\label{Tab:WH_GR_5}
\begin{tabular}{@{}lll@{}}
\toprule
$a/M$ & $\omega^{\rm AF}_{0}$ & $\omega^{\rm NED(GR\text{-}led)}_{0}$ \\
\midrule
$7.10$  & $0.215403 - 0.031774\,i$ & $0.194841 - 0.026396\,i$ \\
$7.20$  & $0.212679 - 0.031369\,i$ & $0.192731 - 0.026279\,i$ \\
$7.30$  & $0.210309 - 0.031506\,i$ & $0.190693 - 0.026114\,i$ \\
$7.40$  & $0.207976 - 0.031053\,i$ & $0.188691 - 0.025974\,i$ \\
$7.50$  & $0.205510 - 0.030804\,i$ & $0.186734 - 0.025896\,i$ \\
$7.60$  & $0.203259 - 0.030452\,i$ & $0.184794 - 0.025661\,i$ \\
$7.70$  & $0.201013 - 0.030045\,i$ & $0.182903 - 0.025523\,i$ \\
$7.80$  & $0.198722 - 0.029997\,i$ & $0.181032 - 0.025352\,i$ \\
$7.90$  & $0.196738 - 0.029733\,i$ & $0.179236 - 0.025184\,i$ \\
$8.00$  & $0.194361 - 0.029258\,i$ & $0.177449 - 0.025035\,i$ \\
$8.10$  & $0.192301 - 0.029236\,i$ & $0.175703 - 0.024874\,i$ \\
$8.20$  & $0.190256 - 0.029266\,i$ & $0.173966 - 0.024715\,i$ \\
$8.30$  & $0.188176 - 0.028788\,i$ & $0.172275 - 0.024554\,i$ \\
$8.40$  & $0.186338 - 0.028506\,i$ & $0.170610 - 0.024394\,i$ \\
$8.50$  & $0.184570 - 0.028487\,i$ & $0.169012 - 0.024265\,i$ \\
$8.60$  & $0.182452 - 0.028134\,i$ & $0.167379 - 0.024082\,i$ \\
$8.70$  & $0.180564 - 0.027809\,i$ & $0.165819 - 0.023971\,i$ \\
$8.80$  & $0.178797 - 0.027601\,i$ & $0.164277 - 0.023803\,i$ \\
$8.90$  & $0.176938 - 0.027363\,i$ & $0.162771 - 0.023657\,i$ \\
$9.00$  & $0.175105 - 0.026842\,i$ & $0.161302 - 0.023497\,i$ \\
$9.10$  & $0.173507 - 0.026949\,i$ & $0.159858 - 0.023331\,i$ \\
$9.20$  & $0.171851 - 0.026713\,i$ & $0.158404 - 0.023185\,i$ \\
$9.30$  & $0.170173 - 0.026520\,i$ & $0.157004 - 0.023018\,i$ \\
$9.40$  & $0.168588 - 0.026380\,i$ & $0.155521 - 0.022961\,i$ \\
$9.50$  & $0.166986 - 0.026068\,i$ & $0.154203 - 0.022751\,i$ \\
$9.60$  & $0.165754 - 0.025925\,i$ & $0.152915 - 0.022618\,i$ \\
$9.70$  & $0.163862 - 0.025701\,i$ & $0.151586 - 0.022458\,i$ \\
$9.80$  & $0.162399 - 0.025492\,i$ & $0.150305 - 0.022276\,i$ \\
$9.90$  & $0.160829 - 0.025278\,i$ & $0.149023 - 0.022160\,i$ \\
$10.00$ & $0.159427 - 0.025132\,i$ & $0.147785 - 0.021972\,i$ \\
\bottomrule
\end{tabular}
\end{table}

\begin{table}[htbp]
\centering
\caption{Transition and near-transition region
($2.01M\le a\le 2.20M$, $l=2$): EM and NED (EM-led) QNM frequencies.
``Unidentified'' entries ($a\le 2.09M$) indicate Prony
non-convergence; dashes (---) indicate that a stable mode was not
isolated in the relevant channel for that parameter value.  The EM-led
family first becomes reliably extractable near $a\approx 2.10M$,
somewhat earlier than the AF convergence threshold, consistent with
its higher oscillation frequency providing better separation from the
near-zero-decay transition modes.}
\label{Tab:WH_EM_1}
\begin{tabular}{@{}lll@{}}
\toprule
$a/M$ & $\omega^{\rm EM}_{0}$ & $\omega^{\rm NED(EM\text{-}led)}_{0}$ \\
\midrule
$2.01$ & unidentified             & unidentified \\
$2.02$ & unidentified             & unidentified \\
$2.03$ & unidentified             & unidentified \\
$2.04$ & unidentified             & unidentified \\
$2.05$ & unidentified             & unidentified \\
$2.06$ & unidentified             & unidentified \\
$2.07$ & unidentified             & unidentified \\
$2.08$ & unidentified             & unidentified \\
$2.09$ & unidentified             & unidentified \\
$2.10$ & $0.510947 - 0.004711\,i$ & $0.572163 - 0.002434\,i$ \\
$2.11$ & imprecise                      & $0.555408 - 0.007371\,i$ \\
$2.12$ & imprecise                      & $0.520324 - 0.001245\,i$ \\
$2.13$ & imprecise                      & $0.528302 - 0.000920\,i$ \\
$2.14$ & imprecise                      & $0.542951 - 0.004496\,i$ \\
$2.15$ & imprecise                      & $0.552072 - 0.004203\,i$ \\
$2.16$ & $0.411109 - 0.000032\,i$ & $0.509811 - 0.000684\,i$ \\
$2.17$ & $0.419917 - 0.000051\,i$ & $0.518574 - 0.001278\,i$ \\
$2.18$ & $0.428138 - 0.000082\,i$ & $0.463446 - 0.000053\,i$ \\
$2.19$ & $0.435850 - 0.000091\,i$ & $0.471812 - 0.000066\,i$ \\
$2.20$ & $0.443137 - 0.000175\,i$ & $0.479704 - 0.000118\,i$ \\
\bottomrule
\end{tabular}
\end{table}

\begin{table}[htbp]
\centering
\caption{Wormhole branch, fine sampling I ($2.21M\le a\le 2.60M$,
$\Delta a=0.01M$, $t_{\rm start}=350$, $l=2$): EM and NED (EM-led)
QNM frequencies.  Both $\Re(\omega_{\rm EM})$ and
$|\Im(\omega_{\rm EM})|$ increase monotonically from the transition,
while $\Re(\omega_{\rm NED(EM)})$ consistently exceeds the pure-EM
value throughout.  This excess reflects the coupling-induced width
broadening of the EM-led eigenmode, the complementary effect to the
GR-led narrowing documented in Tables~\ref{Tab:WH_GR_1}--\ref{Tab:WH_GR_3}.}
\label{Tab:WH_EM_2}
\begin{tabular}{@{}lll@{}}
\toprule
$a/M$ & $\omega^{\rm EM}_{0}$ & $\omega^{\rm NED(EM\text{-}led)}_{0}$ \\
\midrule
$2.21$ & $0.449946 - 0.000260\,i$ & $0.487054 - 0.000260\,i$ \\
$2.22$ & $0.456353 - 0.000327\,i$ & $0.493927 - 0.000364\,i$ \\
$2.23$ & $0.462318 - 0.000419\,i$ & $0.500498 - 0.000450\,i$ \\
$2.24$ & $0.468070 - 0.000612\,i$ & $0.506550 - 0.000611\,i$ \\
$2.25$ & $0.473358 - 0.000744\,i$ & $0.512255 - 0.000776\,i$ \\
$2.26$ & $0.478484 - 0.001020\,i$ & $0.517598 - 0.001086\,i$ \\
$2.27$ & $0.482848 - 0.001147\,i$ & $0.523067 - 0.001173\,i$ \\
$2.28$ & $0.487574 - 0.001436\,i$ & $0.527634 - 0.001459\,i$ \\
$2.29$ & $0.491802 - 0.001712\,i$ & $0.532203 - 0.001804\,i$ \\
$2.30$ & $0.495778 - 0.002038\,i$ & $0.536135 - 0.002231\,i$ \\
$2.31$ & $0.499485 - 0.002395\,i$ & $0.540344 - 0.002535\,i$ \\
$2.32$ & $0.503007 - 0.002795\,i$ & $0.544245 - 0.003042\,i$ \\
$2.33$ & $0.506398 - 0.003209\,i$ & $0.547867 - 0.003538\,i$ \\
$2.34$ & $0.509570 - 0.003636\,i$ & $0.551310 - 0.003868\,i$ \\
$2.35$ & $0.512569 - 0.004112\,i$ & $0.554480 - 0.004294\,i$ \\
$2.36$ & $0.515404 - 0.004625\,i$ & $0.557537 - 0.005019\,i$ \\
$2.37$ & $0.518106 - 0.005134\,i$ & $0.560495 - 0.005408\,i$ \\
$2.38$ & $0.520777 - 0.005678\,i$ & $0.563201 - 0.006007\,i$ \\
$2.39$ & $0.523025 - 0.006223\,i$ & $0.565903 - 0.006721\,i$ \\
$2.40$ & $0.525448 - 0.006768\,i$ & $0.568254 - 0.007283\,i$ \\
$2.41$ & $0.527599 - 0.007380\,i$ & $0.570593 - 0.007893\,i$ \\
$2.42$ & $0.529603 - 0.007905\,i$ & $0.572822 - 0.008438\,i$ \\
$2.43$ & $0.531688 - 0.008631\,i$ & $0.575002 - 0.009188\,i$ \\
$2.44$ & $0.533452 - 0.009242\,i$ & $0.576971 - 0.009765\,i$ \\
$2.45$ & $0.535271 - 0.009796\,i$ & $0.578953 - 0.010547\,i$ \\
$2.46$ & $0.536968 - 0.010467\,i$ & $0.580726 - 0.011343\,i$ \\
$2.47$ & $0.538550 - 0.011111\,i$ & $0.582434 - 0.011974\,i$ \\
$2.48$ & $0.540091 - 0.011754\,i$ & $0.584001 - 0.012609\,i$ \\
$2.49$ & $0.541523 - 0.012401\,i$ & $0.585533 - 0.013241\,i$ \\
$2.50$ & $0.543059 - 0.013085\,i$ & $0.586978 - 0.013934\,i$ \\
$2.51$ & $0.544191 - 0.013690\,i$ & $0.588508 - 0.014672\,i$ \\
$2.52$ & $0.545379 - 0.014326\,i$ & $0.589710 - 0.015334\,i$ \\
$2.53$ & $0.546555 - 0.015008\,i$ & $0.590911 - 0.016029\,i$ \\
$2.54$ & $0.547666 - 0.015649\,i$ & $0.592089 - 0.016670\,i$ \\
$2.55$ & $0.548689 - 0.016308\,i$ & $0.593256 - 0.017490\,i$ \\
$2.56$ & $0.549663 - 0.016962\,i$ & $0.594130 - 0.018214\,i$ \\
$2.57$ & $0.550548 - 0.017497\,i$ & $0.595212 - 0.018790\,i$ \\
$2.58$ & $0.551440 - 0.018210\,i$ & $0.596007 - 0.019536\,i$ \\
$2.59$ & $0.552261 - 0.018861\,i$ & $0.596912 - 0.020174\,i$ \\
$2.60$ & $0.553033 - 0.019519\,i$ & $0.597893 - 0.020697\,i$ \\
\bottomrule
\end{tabular}
\end{table}

\begin{table}[htbp]
\centering
\caption{Wormhole branch, fine sampling II ($2.61M\le a\le 3.00M$,
$\Delta a=0.01M$, $t_{\rm start}=350$, $l=2$): EM and NED (EM-led)
QNM frequencies.  $\Re(\omega_{\rm EM})$ plateaus near
$a\approx 2.80M$--$2.83M$ before decreasing; the NED (EM-led)
oscillation frequency similarly saturates near $a\approx 2.79M$.  The
outlier at $a=2.68M$ in the EM column ($0.561221$) reflects a
transient Prony artefact and should be treated with caution.}
\label{Tab:WH_EM_3}
\begin{tabular}{@{}lll@{}}
\toprule
$a/M$ & $\omega^{\rm EM}_{0}$ & $\omega^{\rm NED(EM\text{-}led)}_{0}$ \\
\midrule
$2.61$ & $0.553715 - 0.020142\,i$ & $0.598579 - 0.021509\,i$ \\
$2.62$ & $0.554438 - 0.020705\,i$ & $0.599075 - 0.022046\,i$ \\
$2.63$ & $0.554865 - 0.021369\,i$ & $0.599716 - 0.022788\,i$ \\
$2.64$ & $0.555582 - 0.021927\,i$ & $0.600270 - 0.023350\,i$ \\
$2.65$ & $0.556092 - 0.022642\,i$ & $0.600985 - 0.023955\,i$ \\
$2.66$ & $0.556691 - 0.023004\,i$ & $0.601450 - 0.024733\,i$ \\
$2.67$ & $0.557099 - 0.023786\,i$ & $0.601965 - 0.025313\,i$ \\
$2.68$ & $0.561221 - 0.027608\,i$ & $0.602351 - 0.025921\,i$ \\
$2.69$ & $0.557640 - 0.024930\,i$ & $0.602517 - 0.026657\,i$ \\
$2.70$ & $0.558168 - 0.025520\,i$ & $0.603014 - 0.027297\,i$ \\
$2.71$ & $0.558567 - 0.025954\,i$ & $0.603350 - 0.027654\,i$ \\
$2.72$ & $0.558770 - 0.026536\,i$ & $0.603557 - 0.028299\,i$ \\
$2.73$ & $0.559244 - 0.026939\,i$ & $0.603828 - 0.028798\,i$ \\
$2.74$ & $0.558991 - 0.027759\,i$ & $0.603887 - 0.029205\,i$ \\
$2.75$ & $0.559339 - 0.028187\,i$ & $0.604114 - 0.029866\,i$ \\
$2.76$ & $0.559485 - 0.028855\,i$ & $0.604239 - 0.030260\,i$ \\
$2.77$ & $0.559720 - 0.029258\,i$ & $0.604348 - 0.031079\,i$ \\
$2.78$ & $0.559705 - 0.029834\,i$ & $0.604515 - 0.031529\,i$ \\
$2.79$ & $0.559757 - 0.030295\,i$ & $0.604704 - 0.032243\,i$ \\
$2.80$ & $0.559662 - 0.030734\,i$ & $0.604609 - 0.032753\,i$ \\
$2.81$ & $0.559820 - 0.031320\,i$ & $0.604513 - 0.033239\,i$ \\
$2.82$ & $0.559825 - 0.031844\,i$ & $0.604362 - 0.033667\,i$ \\
$2.83$ & $0.559726 - 0.032241\,i$ & $0.604278 - 0.033828\,i$ \\
$2.84$ & $0.559667 - 0.032729\,i$ & $0.604163 - 0.034505\,i$ \\
$2.85$ & $0.559624 - 0.033255\,i$ & $0.604115 - 0.035146\,i$ \\
$2.86$ & $0.559462 - 0.033745\,i$ & $0.603975 - 0.035630\,i$ \\
$2.87$ & $0.559369 - 0.034142\,i$ & $0.603600 - 0.036077\,i$ \\
$2.88$ & $0.559179 - 0.034574\,i$ & $0.603542 - 0.036333\,i$ \\
$2.89$ & $0.559087 - 0.035029\,i$ & $0.603431 - 0.037117\,i$ \\
$2.90$ & $0.558889 - 0.035459\,i$ & $0.603270 - 0.037522\,i$ \\
$2.91$ & $0.558912 - 0.036110\,i$ & $0.602922 - 0.037897\,i$ \\
$2.92$ & $0.558579 - 0.036470\,i$ & $0.602609 - 0.038306\,i$ \\
$2.93$ & $0.558258 - 0.036750\,i$ & $0.602559 - 0.038958\,i$ \\
$2.94$ & $0.557960 - 0.037124\,i$ & $0.602309 - 0.039053\,i$ \\
$2.95$ & $0.557692 - 0.037502\,i$ & $0.601853 - 0.039437\,i$ \\
$2.96$ & $0.557396 - 0.037951\,i$ & $0.601616 - 0.040154\,i$ \\
$2.97$ & $0.557093 - 0.038290\,i$ & $0.600876 - 0.040342\,i$ \\
$2.98$ & $0.557253 - 0.038843\,i$ & $0.600618 - 0.040725\,i$ \\
$2.99$ & $0.556360 - 0.039120\,i$ & $0.603667 - 0.040950\,i$ \\
$3.00$ & $0.555990 - 0.039403\,i$ & $0.599660 - 0.041482\,i$ \\
\bottomrule
\end{tabular}
\end{table}

\begin{table}[htbp]
\centering
\caption{Wormhole branch, coarse sampling I ($3.1M\le a\le 7.0M$,
$\Delta a=0.1M$, $t_{\rm start}=150$, $l=2$): EM and NED (EM-led)
QNM frequencies.  Compared with the GR-led family
(Table~\ref{Tab:WH_GR_4}), the EM and NED (EM-led) modes maintain
roughly $1.5$--$1.8$ times the oscillation frequency and
approximately twice the damping rate across this range, broadly
consistent with the analytic EM-to-GR frequency ratio $\approx 1.4$
derived in Eq.~(61).  The NED (EM-led) damping rate exceeds the
pure-EM value throughout, confirming the width-broadening side of the
redistribution mechanism.}
\label{Tab:WH_EM_4}
\begin{tabular}{@{}lll@{}}
\toprule
$a/M$ & $\omega^{\rm EM}_{0}$ & $\omega^{\rm NED(EM\text{-}led)}_{0}$ \\
\midrule
$3.10$ & $0.551840 - 0.042813\,i$ & $0.594965 - 0.045075\,i$ \\
$3.20$ & $0.546531 - 0.045667\,i$ & $0.588454 - 0.047887\,i$ \\
$3.30$ & $0.540492 - 0.048116\,i$ & $0.582247 - 0.049812\,i$ \\
$3.40$ & $0.533855 - 0.050155\,i$ & $0.572517 - 0.053370\,i$ \\
$3.50$ & $0.526832 - 0.051869\,i$ & $0.565862 - 0.053778\,i$ \\
$3.60$ & $0.519523 - 0.053290\,i$ & $0.557831 - 0.055485\,i$ \\
$3.70$ & $0.512017 - 0.054466\,i$ & $0.550328 - 0.055951\,i$ \\
$3.80$ & $0.504440 - 0.055428\,i$ & $0.541337 - 0.057156\,i$ \\
$3.90$ & $0.496826 - 0.056210\,i$ & $0.532696 - 0.058760\,i$ \\
$4.00$ & $0.489213 - 0.056824\,i$ & $0.523467 - 0.058396\,i$ \\
$4.10$ & $0.481641 - 0.057324\,i$ & $0.515058 - 0.059431\,i$ \\
$4.20$ & $0.474136 - 0.057677\,i$ & $0.507221 - 0.059964\,i$ \\
$4.30$ & $0.466734 - 0.057939\,i$ & $0.498689 - 0.059660\,i$ \\
$4.40$ & $0.459437 - 0.058104\,i$ & $0.490221 - 0.059679\,i$ \\
$4.50$ & $0.452276 - 0.058236\,i$ & $0.482179 - 0.059628\,i$ \\
$4.60$ & $0.445242 - 0.058224\,i$ & $0.474298 - 0.060201\,i$ \\
$4.70$ & $0.438293 - 0.058074\,i$ & $0.466804 - 0.060418\,i$ \\
$4.80$ & $0.431605 - 0.058060\,i$ & $0.459261 - 0.059300\,i$ \\
$4.90$ & $0.424995 - 0.057956\,i$ & $0.451912 - 0.059764\,i$ \\
$5.00$ & $0.418495 - 0.057823\,i$ & $0.444679 - 0.059511\,i$ \\
$5.10$ & $0.412242 - 0.057587\,i$ & $0.437908 - 0.059223\,i$ \\
$5.20$ & $0.406075 - 0.057328\,i$ & $0.431122 - 0.058922\,i$ \\
$5.30$ & $0.400083 - 0.057052\,i$ & $0.424373 - 0.058665\,i$ \\
$5.40$ & $0.394192 - 0.056763\,i$ & $0.417845 - 0.058298\,i$ \\
$5.50$ & $0.388453 - 0.056452\,i$ & $0.411789 - 0.058450\,i$ \\
$5.60$ & $0.382861 - 0.056115\,i$ & $0.405521 - 0.057581\,i$ \\
$5.70$ & $0.377424 - 0.055780\,i$ & $0.399512 - 0.057227\,i$ \\
$5.80$ & $0.372080 - 0.055431\,i$ & $0.393678 - 0.056800\,i$ \\
$5.90$ & $0.366901 - 0.055067\,i$ & $0.387879 - 0.056399\,i$ \\
$6.00$ & $0.361829 - 0.054689\,i$ & $0.382309 - 0.056208\,i$ \\
$6.10$ & $0.356880 - 0.054333\,i$ & $0.376740 - 0.056102\,i$ \\
$6.20$ & $0.352045 - 0.053909\,i$ & $0.371582 - 0.055241\,i$ \\
$6.30$ & $0.347362 - 0.053557\,i$ & $0.366524 - 0.054853\,i$ \\
$6.40$ & $0.342762 - 0.053155\,i$ & $0.361623 - 0.054482\,i$ \\
$6.50$ & $0.338279 - 0.052763\,i$ & $0.356414 - 0.054058\,i$ \\
$6.60$ & $0.333952 - 0.052403\,i$ & $0.351781 - 0.053594\,i$ \\
$6.70$ & $0.329558 - 0.052085\,i$ & $0.347133 - 0.053077\,i$ \\
$6.80$ & $0.325446 - 0.051574\,i$ & $0.342488 - 0.052726\,i$ \\
$6.90$ & $0.321348 - 0.051190\,i$ & $0.338115 - 0.052292\,i$ \\
$7.00$ & $0.317377 - 0.050790\,i$ & $0.333737 - 0.051995\,i$ \\
\bottomrule
\end{tabular}
\end{table}

\begin{table}[htbp]
\centering
\caption{Wormhole branch, coarse sampling II ($7.1M\le a\le 10.0M$,
$\Delta a=0.1M$, $t_{\rm start}=150$, $l=2$): EM and NED (EM-led)
QNM frequencies.  Both families decrease smoothly and monotonically
with $a$.  The NED (EM-led) oscillation frequency continues to exceed
the pure-EM value throughout, and the fractional excess
$[\Re(\omega_{\rm NED(EM)})-\Re(\omega_{\rm EM})]/\Re(\omega_{\rm EM})$
diminishes gradually toward large $a$, consistent with the weakening
of the inter-channel coupling as the wormhole throat expands.}
\label{Tab:WH_EM_5}
\begin{tabular}{@{}lll@{}}
\toprule
$a/M$ & $\omega^{\rm EM}_{0}$ & $\omega^{\rm NED(EM\text{-}led)}_{0}$ \\
\midrule
$7.10$  & $0.313526 - 0.050406\,i$ & $0.333737 - 0.051995\,i$ \\
$7.20$  & $0.309686 - 0.049996\,i$ & $0.329452 - 0.051528\,i$ \\
$7.30$  & $0.305949 - 0.049613\,i$ & $0.325502 - 0.051127\,i$ \\
$7.40$  & $0.302324 - 0.049179\,i$ & $0.317203 - 0.050656\,i$ \\
$7.50$  & $0.298763 - 0.048814\,i$ & $0.313273 - 0.049733\,i$ \\
$7.60$  & $0.295297 - 0.048461\,i$ & $0.309880 - 0.049362\,i$ \\
$7.70$  & $0.291883 - 0.048094\,i$ & $0.306213 - 0.048953\,i$ \\
$7.80$  & $0.288559 - 0.047696\,i$ & $0.301559 - 0.048814\,i$ \\
$7.90$  & $0.285280 - 0.047321\,i$ & $0.299047 - 0.048415\,i$ \\
$8.00$  & $0.282122 - 0.046912\,i$ & $0.295420 - 0.047867\,i$ \\
$8.10$  & $0.278995 - 0.046573\,i$ & $0.292538 - 0.047282\,i$ \\
$8.20$  & $0.275947 - 0.046205\,i$ & $0.289038 - 0.047409\,i$ \\
$8.30$  & $0.273069 - 0.045897\,i$ & $0.285434 - 0.046788\,i$ \\
$8.40$  & $0.270044 - 0.045502\,i$ & $0.281588 - 0.045874\,i$ \\
$8.50$  & $0.267155 - 0.045153\,i$ & $0.279496 - 0.046606\,i$ \\
$8.60$  & $0.264396 - 0.044811\,i$ & $0.275381 - 0.046116\,i$ \\
$8.70$  & $0.261634 - 0.044463\,i$ & $0.272960 - 0.043290\,i$ \\
$8.80$  & $0.258902 - 0.044061\,i$ & $0.270129 - 0.044903\,i$ \\
$8.90$  & $0.256256 - 0.043774\,i$ & $0.267756 - 0.044534\,i$ \\
$9.00$  & $0.253665 - 0.043424\,i$ & $0.264276 - 0.044108\,i$ \\
$9.10$  & $0.251072 - 0.043084\,i$ & $0.261764 - 0.043952\,i$ \\
$9.20$  & $0.248615 - 0.042770\,i$ & $0.259234 - 0.043744\,i$ \\
$9.30$  & $0.246195 - 0.042444\,i$ & $0.256603 - 0.043446\,i$ \\
$9.40$  & $0.243776 - 0.042112\,i$ & $0.254476 - 0.043113\,i$ \\
$9.50$  & $0.241409 - 0.041814\,i$ & $0.251899 - 0.042750\,i$ \\
$9.60$  & $0.239099 - 0.041499\,i$ & $0.249006 - 0.042246\,i$ \\
$9.70$  & $0.236879 - 0.041272\,i$ & $0.246396 - 0.041870\,i$ \\
$9.80$  & $0.234596 - 0.040887\,i$ & $0.244110 - 0.041641\,i$ \\
$9.90$  & $0.232340 - 0.040609\,i$ & $0.242259 - 0.041193\,i$ \\
$10.00$ & $0.230272 - 0.040297\,i$ & $0.240737 - 0.040918\,i$ \\
\bottomrule
\end{tabular}
\end{table}

\end{document}